\documentclass[twocolumn]{aastex63}
\usepackage{xcolor}
\begin{document}

\title{Fast Blue Optical Transients due to Circumstellar Interaction and the Mysterious Supernova SN 2018gep}

%% Notice placement of commas and superscripts and use of &
%% in the author list

\shorttitle{SN 2018gep: Dual intrepretations}
\shortauthors{Leung, Fuller and Nomoto}

%[0000-0002-4972-3803]
\author[0000-0002-4972-3803]{Shing-Chi Leung}
\affiliation{TAPIR, Walter Burke Institute for Theoretical Physics, 
Mailcode 350-17, Caltech, Pasadena, CA 91125, USA}
%\email{scleung@caltech.edu}

%[0000-0002-4544-0750]
\author[0000-0002-4544-0750]{Jim Fuller}
%\thanks{Email address: jfuller@caltech.edu}

\affiliation{TAPIR, Walter Burke Institute for Theoretical Physics, 
Mailcode 350-17, Caltech, Pasadena, CA 91125, USA}

%[0000-0001-9553-0685]
\author[0000-0001-9553-0685]{Ken'ichi Nomoto}
%\thanks{Email address: nomoto@astron.s.u-tokyo.ac.jp}

\affiliation{Kavli Institute for the Physics and 
Mathematics of the Universe (WPI), The University 
of Tokyo Institutes for Advanced Study, The 
University of Tokyo, Kashiwa, Chiba 277-8583, Japan}

%\author{Others}

\correspondingauthor{Shing-Chi Leung}
\email{scleung@caltech.edu}

\received{10 March 2021}  
\revised{26 April 2021}
\accepted{27 April 2021}   
\submitjournal{Astrophysical Journal}

\date{\today}

\begin{abstract}

The discovery of SN 2018gep (ZTF18abukavn) challenged our understanding of the late-phase evolution of massive stars and their supernovae (SNe).
%after the mysterious rapid transient At2018cow (ATLAS18qqn).
The fast rise in luminosity of this SN (spectroscopically classified as a broad-lined Type Ic SN), indicates that the ejecta interacts with a dense circumstellar medium (CSM), while an additional energy source such as $^{56}$Ni-decay is required to explain the late-time light curve. These features hint at the explosion of a massive star with pre-supernova mass-loss. In this work, we examine the physical origins of rapidly evolving astrophysical transients like SN 2018gep. We investigate the wave-driven mass-loss mechanism and how it depends on model parameters such as progenitor mass and deposition energy, searching for stellar progenitor models that can reproduce the observational data. A model with an ejecta mass $\sim \! 2 \, M_{\odot}$, explosion energy $\sim \! 10^{52}$ erg, a circumstellar medium of mass $\sim \! 0.3 \, M_{\odot}$ and radius $\sim \! 1000 \, R_{\odot}$, and a $^{56}$Ni mass of $\sim \! 0.3 \, M_{\odot}$ provides a good fit to the bolometric light curve. We also examine how interaction-powered light curves depend more generally on these parameters, and how ejecta velocities can help break degeneracies. We find both wave-driven mass-loss and mass ejection via pulsational pair-instability can plausibly create the dense CSM in SN 2018gep, but we favor the latter possibility.

\end{abstract}



\section{Introduction}

\subsection{Circumstellar Medium Interaction Mechanism}

Interaction between the supernova (SN) ejecta and dense circumstellar material (CSM) \citep{Chevalier1982} is one of the proposed models \citep[e.g.,][]{Woosley2007, Moriya2013a, Chatzopoulos2013, Morozova2015, Sorokina2016, Blinnikov2017, Kasen2017, Smith2017, Moriya2018b} to explain the diversity of highly luminous supernovae (SNe), in parallel with magnetar-powered SNe \citep{Maeda2007, Woosley2010, Kasen2010, Kasen2016},
accretion-powered SNe \citep{Dexter2013, Wang2018} and 
pair-instability SNe \citep{Kasen2011, Gilmer2017}.
CSM interaction occurs when the rapidly expanding stellar ejecta collides with the quasi-static CSM. The ejecta drives a shock through the CSM, which converts the ejecta kinetic energy into thermal energy, leading to the bright event observed. The model has been applied to recent super-luminous SNe such as SN 2008gy \citep{Woosley2007}, 
SN 2007bi, SN 2010gx and PTF09cnd \citep{Sorokina2016},
iPTF14hls \citep{Woosley2018}, PTF12dam \citep{Tolstov2017},
SN 2016iet \citep{Lunnan2018}, iPTF16eh \citep{Gomez2019},
AT2018cow \citep{Leung2020COW} and PS15dpn \citep{Wang2020}.

To explain the origin of the CSM, a number of mechanisms can trigger greatly enhanced mass loss prior to the final stellar explosion, including common envelope-triggered mass loss \citep{Chevalier2012, Schroeder2020}, pulsation-induced mass-loss in pulsational pair-instability supernovae (PPISNe) \citep{Umeda2003, Woosley2017, Marchant2019, Leung2019PPISN, Leung2018PPISN2, Woosley2019, Renzo2020}, enhanced stellar wind in super-AGB stars \citep{Jones2013, Moriya2014, Nomoto2017, Tolstov2019, Leung2019PASA} and wave-driven mass-loss \citep{Quataert2012, Shiode2014, Fuller2017, Fuller2018, Ryoma2019, Leung2020Wave, Kuriyama2020}.

Pulsation-induced mass loss relies on the electron-positron pair-creation catastrophe \citep{Barkat1967} which happens in very massive stars ($\sim \! 80 - 140 \, M_{\odot}$) \citep{Umeda2002, Heger2002, Ohkubo2009, Hirschi2017,Yoshida2016}.
These radiation-dominated stars lose core pressure support from photons during their conversion into electron-positron pairs.
The contraction of the star's core becomes dynamical,
triggering explosive burning of C and O. The excess energy generation makes the core bounce, driving a shock through the envelope that ejects mass from the surface. The star can experience several mass-loss events, depending on the available carbon and oxygen in the core \citep{Woosley2017, Marchant2019, Leung2019PPISN, Woosley2019, Renzo2020}. 

In \cite{Leung2020COW}, we explored the possibility of applying the pulsation-induced mass-loss model to explain the rapid transient AT2018cow. Like SN 2018gep, AT2018cow is also classified as a ``Fast Blue Optical Transient" (FBOT). That work, together with \cite{Tolstov2017}, suggests that pulsational pair-instability supernovae (PPISNe) provides the flexibility to span the wide diversity of transient objects from FBOT to super-luminous SNe. However, the unusually rapid transient in SN 2018gep leads to speculation whether other mass loss mechanisms are necessary to explain the optical signals of this object. 

Wave-driven mass loss \citep{Quataert2012, Shiode2014, Fuller2017, Fuller2018,Wu2020}
relies on the vigorous convective motions in the massive star's core during its late-phase nuclear burning (e.g., carbon-burning and later advanced burning stages).
This in turns excites internal gravity waves that propagate through the radiative core, where some of the wave energy is transmitted into the envelope via acoustic waves. When the waves reach the surface where the density gradient is large, they develop into weak shocks that dissipate and deposit their energy in the surroundings. Even though only a small fraction of the wave energy is leaked to the envelope,
it can be sufficient to eject a substantial amount of mass.

\subsection{SN2018gep as a Rapid Transient}

SN 2018gep (ZTF18abukavn) is a supernova discovered by the Zwicky Transient Factory (ZTF) \citep{Bellm2019a, Graham2019}, first analyzed by \cite{Ho2019} and later by \citep{Pritchard2020}.
%at a redshift $z = 0.03154$.
This object has photometric and spectroscopic features similar to some other recent rapid transients, such as AT2018cow \citep{Smartt2018, Prentice2018, Perley2019, Margutti2019} and iPTF16asu \citep{Wang2019,Whitesides2017}. SN2018gep is remarkable for its very early detection by ZTF and its proximity at $\sim143$ Mpc from the Earth, which allowed for detailed follow-up observations.

Among all rapid transients, SN 2018gep has a fast rise time of 0.5 -- 3 days \citep{Ho2019} until it reached a large peak luminosity of $3 \times 10^{44}$ erg s$^{-1}$. Its low-mass and compact host galaxy has a metallicity only one-fifth of solar metallicity. The surface temperature of the transient at its peak of $\sim40000$ K is the among the highest observed in stripped-envelope SNe. The upper limits of detection by high energy bands (X-ray and gamma-ray) and the radio band have led to speculations that this explosion arises from interaction with a compact and dense CSM. Future polarization and X-ray observations of this object can further pin down the possibility of dense CSM \citep[see, e.g.,][for applications in AT2018cow.]{Huang2019, Rivera2019}

\subsection{Outline}

In Section \ref{sec:Progenitor}, we perform hydrodynamical modeling of mass loss from helium star SN progenitors due to wave heating, predicting the resulting CSM structure at the time of explosion.
In Section \ref{sec:interpret} we compare the CSM properties from our wave-driven mass loss models to radiative transfer models of this work and those in \cite{Ho2019}. We also include PPISN models from the literature to contrast the corresponding CSM properties of this class of stars. 
In Section \ref{sec:radtran} we present radiative transfer modelling and an explosion parameter search to fit the bolometric light curve of SN 2018gep.
We then discuss the robustness of our results in Section \ref{sec:discussion}, examining possible degeneracies, comparing to other models of SN 2018gep, and discussing the physical origins of this event.
Finally we summarize the findings of this work and highlight our interpretation of SN 2018gep.
  
\section{Wave-Driven Mass Loss}
\label{sec:Progenitor}

%Method MESA + SNEC
\subsection{General Modeling Method}
\label{sec:methods}

To investigate wave-driven mass loss and to construct pre-collapse models for radiative transfer modeling, we use the one-dimensional stellar evolution code MESA (Module for Experiments in Stellar Astrophysics) \citep{Paxton2011,Paxton2013,Paxton2015,Paxton2017}, version 8118. We model stars with initial masses $M_{\rm ini} =$ 20 -- 80 $M_{\odot}$, which correspond to He cores with masses $M_{\rm He} \sim$ 5 -- 40 $M_{\odot}$ \citep{Hirschi2017,Woosley2017}. In addition to the default massive star settings provided in MESA, we set the star to be non-rotating, with a Dutch mass loss\footnote{The Dutch mass loss prescription is based on \cite{deJager1988} for cool stars, \cite{Vink2000, Vink2001} for hot hydrogen-rich stars, and \cite{Nugis2000} for Wolf-Rayet stars.}
scaling factor $\eta = 0.8$ and an overshoot parameter $f_{\rm ov} = 0.001$. The final mass before collapse, even without dynamical mass loss, depends strongly on the initial metallicity due to mass loss from line-driven winds. We follow the entire stellar evolution from the onset of core H-burning until the onset of gravitational collapse.\footnote{Configuration files for generating models reported in this article are available at \url{https://doi.org/10.5281/zenodo.4722017}}.

We consider only stripped-envelope stars in this work because this class of stars produce hydrogen-free Type Ib/c SNe like SN 2018gep. We first model a zero-age main-sequence star until hydrogen is exhausted in the core. Then we remove the H-envelope by relaxing the mass to exclude the H-rich matter and continue the evolution. The loss of the majority of the H-envelope is most likely to arise from a binary interaction (either stable or unstable mass transfer). While binary stripping may not remove all of the hydrogen in reality, the remaining hydrogen is quickly lost via winds at the masses and metallicities we consider. The final mass is thus primarily controlled by (uncertain) wind mass loss rather than the details of the binary interaction.

In Table \ref{table:mass}, we list relevant masses for the progenitor models in this study. Although the ZAMS mass is the model parameter we control, we will focus on the pre-explosion mass $M_{\rm exp}$, as it determines the SN ejecta mass, along with the core evolution and hence the wave-driven mass-loss. The core mass is defined by the outer mass coordinate where the abundances of those elements drop below threshold values, where we use the default value 0.01.
%The He core mass grows linearly with the progenitor mass, while the C-core and pre-explosion masses are also linearly increasing with the progenitor mass at a lower rate to a good approximation.

\begin{table}[]
    \centering
    \caption{The initial, He-core, C-core and pre-explosion 
    masses of the models studied in this work. We assume $Z = 0.02$ and a Dutch wind$^1$ coefficient $\eta = 0.8$.
    The He-core mass is taken at the end of core H-burning. The C-core mass is taken from the values at the end of simulations (The definition of "the core mass" is given in the text of Section \ref{sec:methods}.)}
    %All masses are in units of $M_{\odot}$.}
    \begin{tabular}{c|c|c|c}
        \hline 
        $M_{\rm ini}$ ($M_{\odot}$) & $M_{\rm He}$ $(M_{\odot})$ & $M_{\rm C}$ ($M_{\odot}$) & $M_{\rm exp}$ ($M_{\odot}$)\\
        \hline 
        25 & 6.56 & 3.00 & 5.44 \\
        40 & 12.97 & 5.58 & 9.20 \\
        50 & 17.45 & 8.39 & 10.46 \\ 
        60 & 21.78 & 9.58 & 12.22 \\ \hline
    \end{tabular}
    
    \label{table:mass}
\end{table}

\begin{figure*}
\centering
\includegraphics*[width=18cm]{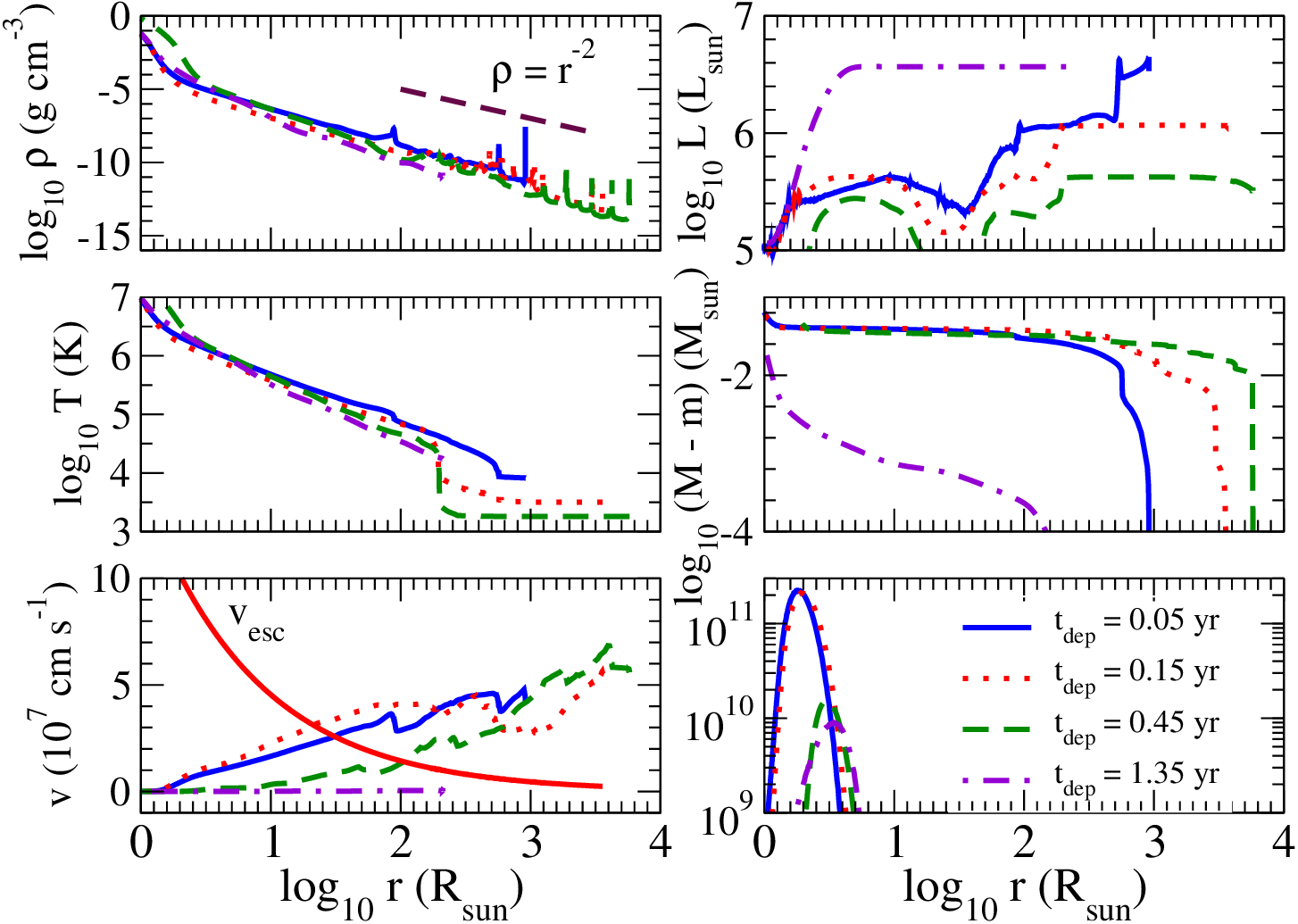}
\caption{
The hydrodynamic profile of stellar models at the end of 
the wave heating simulations including the density (top left), temperature (middle left), velocity (bottom left), luminosity (top right), external mass (middle right) and local energy deposition rate in units of erg s$^{-1}$ (bottom right). The progenitor model is a helium core with a pre-explosion mass $M_{\rm exp} = 5.44~M_{\odot}$ at $Z = 0.02$, and a deposited energy of $6 \times 10^{47}$ erg over a time scale before core-collapse of $t_{\rm dep} = 0.05$ (blue solid line), 0.15 (red dotted line), 0.45 (green dashed line) and 1.35 year (purple dot-dashed line) respectively. In the bottom left plot, we include the escape velocity using the model with $t_{\rm dep} = 0.15$ year. }
\label{fig:profile_tdep}
\end{figure*}

To investigate the process of wave-driven mass loss, we follow a scheme similar to that described in \cite{Fuller2018}. Once an oxygen-burning core has developed in the star, we terminate the simulation. We set the inner boundary at $1 ~M_{\odot}$, interior to the carbon core outer boundary. We keep the outer carbon layer as a buffer, but in general, the mass loss is very small and the carbon layer remains unchanged. 

To calculate how the wave energy is deposited in the envelope, we follow the method of \cite{Fuller2018}, in which the outgoing acoustic wave luminosity $L_{\rm wave}$ decreases as 
\begin{equation}
    \frac{dL_{\rm wave}}{dM} = -\frac{L_{\rm wave}}{M_{\rm wave, shock}} - \frac{L_{\rm wave}}{M_{\rm wave, diff}},
\end{equation}
The local value of $d L_{\rm wave}/dM$ is the deposited energy per unit mass within the star. Above, $M_{\rm wave, shock}$ and $M_{\rm wave, diff}$ are the expected damping mass by shock heating and radiative diffusion, given by:
\begin{eqnarray}
%\begin{align}
    M_{\rm wave, shock} &=& {3 \pi} \frac{\gamma + 1} \omega c_s^2 \left( \frac{L_{\rm max}^3}{L_{\rm wave}} \right)^{1/2}, \\
    M_{\rm wave, diff} &=& \frac{2 L_{\rm max}}{\omega^2 K},
%\end{align}
\end{eqnarray}
where $\gamma$, $c_s^2$ and $K$ are the adiabatic index, sound speed squared and thermal diffusivity \citep{Fuller2017}. $L_{\rm max} = 2 \pi \rho r^2 c_s^3$ is the maximum energy transportable by linear acoustic waves and $\omega$ is the wave frequency. The wave frequencies excited by convection are typically in the range $\omega = 10^{-2} - 10^{-3}$ s$^{-1}$ and are a model parameter in this work.
Additionally, the outflowing matter can expand with a super-sonic velocity, stretching the acoustic wavelength and increasing the value of $M_{\rm wave,diff}$ and $M_{\rm wave,shock}$ as described in \cite{Fuller2018}.

Similar to our previous work \citep{Leung2020Wave}, when we model the envelope dynamics, we treat the duration of energy deposition as a parameter.
This allows us to model the resulting interaction-powered light curve as a function of both the wave energy deposition and its duration, but we note that the two parameters are linked when modeling the inner core evolution of the star, as investigated in \cite{Shiode2014} and \cite{Wu2020}. We retain all of the ejected mass shells, which later form the CSM, on our model grids because the typical timescale from the onset of O-burning to the final collapse is $\sim \! 0.1$ year, and the corresponding CSM radius is less than $10^4~R_{\odot}$.

Below, we find that the mass loss rates via wave energy deposition are much larger than those due to line-driven winds, which are typically $\dot{M}_{\rm wind} \sim 10^{-4} \, M_\odot \, {\rm yr}^{-1}$ near the end of the progenitor's life. Hence, while winds remove much more mass than wave heating during prior phases of evolution, wave-driven mass loss dominates over wind mass loss in the final months of the star's life, so we neglect wind mass loss when constructing CSM density profiles around the progenitor star.

\subsection{Dependence on Deposition Duration}

We first study how the CSM properties depend on the energy deposition duration. We consider a 5.44 $M_{\odot}$ He star with $Z = 0.02$. In the energy deposition phase, we deposit a total of $6 \times 10^{47}$ erg energy in the envelope, a value near the upper end of the envelope wave energy deposition range found in \cite{Wu2020}.
The duration, which depends on the exact duration of O-burning, is treated as a free parameter from 0.05 to 1.35 year, which covers the typical O-burning duration for a wide range of stellar mass. In Figure \ref{fig:profile_tdep}, we plot the hydrodynamical profiles of these models at the end of the simulations.

The density profile shows a generally smooth structure with radial dependence similar to (though slightly steeper than) that for a steady wind, $\rho \propto r^{-2}$. The density profile is also punctuated by spikes associated with internal shocks, which arise due to the uneven energy deposition (and hence uneven outflow velocity) at early times. Near its outer edge, the CSM has a typical density of $10^{-10}$ -- $10^{-13}$ g cm$^{-3}$. The temperature also falls rapidly with radius, except in the outermost optically thin layers where it is constant.
The ejecta moves away from the core with a typical velocity of $\sim 3 - 5 \times 10^7$ cm s$^{-1}$, but with an exception for the longest energy deposition model ($t_{\rm dep} = 1.35$ year), in which the expansion of the star is quasi-hydrostatic. The escape velocity at large radii is about $3 \times 10^6$ cm s$^{-1}$, confirming that the ejected matter can successfully escape from the gravity of the star.  

The luminosity profile demonstrates how the star distributes the deposited wave energy. In the bound portion of the star, the luminosity is constant at $\approx \! 10^5 ~L_{\odot}$. Across the energy deposition zone, the stellar luminosity quickly rises by one to two orders of magnitude, but is smaller than the total wave heating rate. This shows that much of the wave energy is used to unbind material (i.e., increase its gravitational potential energy) and convert it to kinetic energy of the outflow. An exception is the model with $t_{\rm dep}=1.35 \, {\rm yr}$, in which matter is not efficiently ejected and most of the wave energy is radiated from the star. 
A long $t_{\rm dep}$ corresponds to a low $L_{\rm dep} = E_{\rm dep} / t_{\rm dep}$ for a given deposited energy $E_{\rm dep}$. The heated layer has sufficient time to quasi-hydrostatically expand. This prevents the formation of fast winds and/or shocks and hence reduces mass ejection.
Note also that the star's luminosity can decrease or increase within the outflowing material, the latter occurring as kinetic energy from colliding shells is converted into heat that is radiated outward.

The external mass coordinate profile of $M - m$ shows the radial dependence of the ejected mass, with most of the mass located in the outer part of the CSM where the $M-m$ profiles turn over between roughly $300-3000 \, R_{\odot}$. We see that the ejected mass is a few hundredths of a solar mass for all but the longest energy deposition time scale.

We next examine the energy deposition profile when the simulation ends, recalling that the energy deposition zone is time-dependent. The energy deposition zone moves towards to the stellar core as the heated zone expands. We also see that the waves damp at a smaller radius when $t_{\rm dep}$ becomes small, because the corresponding $L_{\rm wave} = E_{\rm dep} / t_{\rm dep}$ becomes large, so that the acoustic waves develop into weak shocks at a higher density, and thus at a smaller radius. The peak of the heat deposition
also occurs near the sonic point of the outflow, i.e., near the boundary between the nearly hydrostatic star and the outflowing CSM. 

\subsection{Dependence on Deposition Energy}

\begin{figure*}
\centering
\includegraphics*[width=18cm]{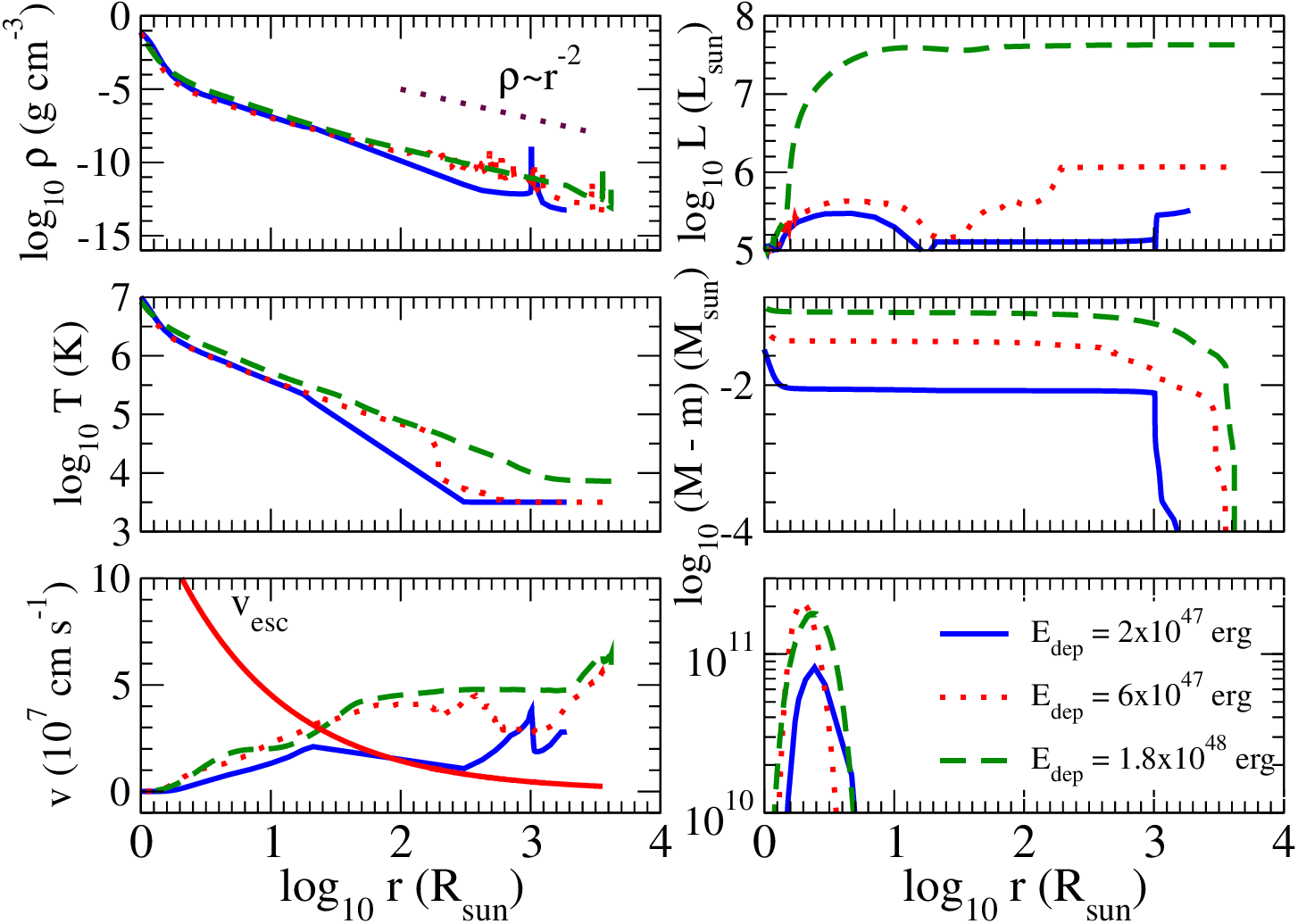}
\caption{
Similar to Figure \ref{fig:profile_tdep} but for models with 
a total energy deposition of $2 \times 10^{47}$, $6 \times 10^{47}$ and $1.8 \times 10^{48}$ erg respectively. A model with a pre-explosion mass of $M_{\rm exp} =$ 5.44 $M_{\odot}$ and a
duration of 0.15 year of energy deposition are used during the simulation. The escape velocity in the bottom left plot used the data from the model with $E_{\rm dep} = 6 \times 10^{47}$ erg.}
\label{fig:profile_energy}
\end{figure*}

We next compare how the total energy deposition changes the CSM profile of the $M_{\rm exp} =$ 5.44 $M_{\odot}$ model, fixing the deposition duration at 0.15 year. In general, the net energy deposition is determined by the core structure and evolution, and \cite{Wu2020} find typical energies of a few $\times 10^{47}$ erg for a wide number of models, with many massive star models in the range $10^{47}-10^{48}$ erg. Hence, we consider total energy depositions of $2 \times 10^{47} \, {\rm erg}$, $6 \times 10^{47} \, {\rm erg}$, and $1.8 \times 10^{48}$. 

Figure \ref{fig:profile_energy} shows that the total energy deposition is another primary factor affecting the CSM. While the density and temperature profiles are fairly similar between the models, the ejecta mass varies by about an order of magnitude, with the model with $E_{\rm dep} = 1.8 \times 10^{48}$ erg ejecting about 0.063 $M_{\odot}$, whereas the model with $2 \times 10^{47}$ erg ejects about 0.009 $M_{\odot}$. The ejected mass scales approximately linearly with the injected energy. The CSM also extends to slightly larger radii for larger energy deposition. The ejecta velocity shows only minor variations between the models, with the low-energy model exhibiting slightly larger velocity at the end of the simulation. The luminosity also does not monotonically change with $E_{\rm dep}$, with a much higher luminosity for the largest $E_{\rm dep}$ model.

\subsection{Dependence on Progenitor Mass}

\begin{figure*}
\centering
\includegraphics*[width=18cm]{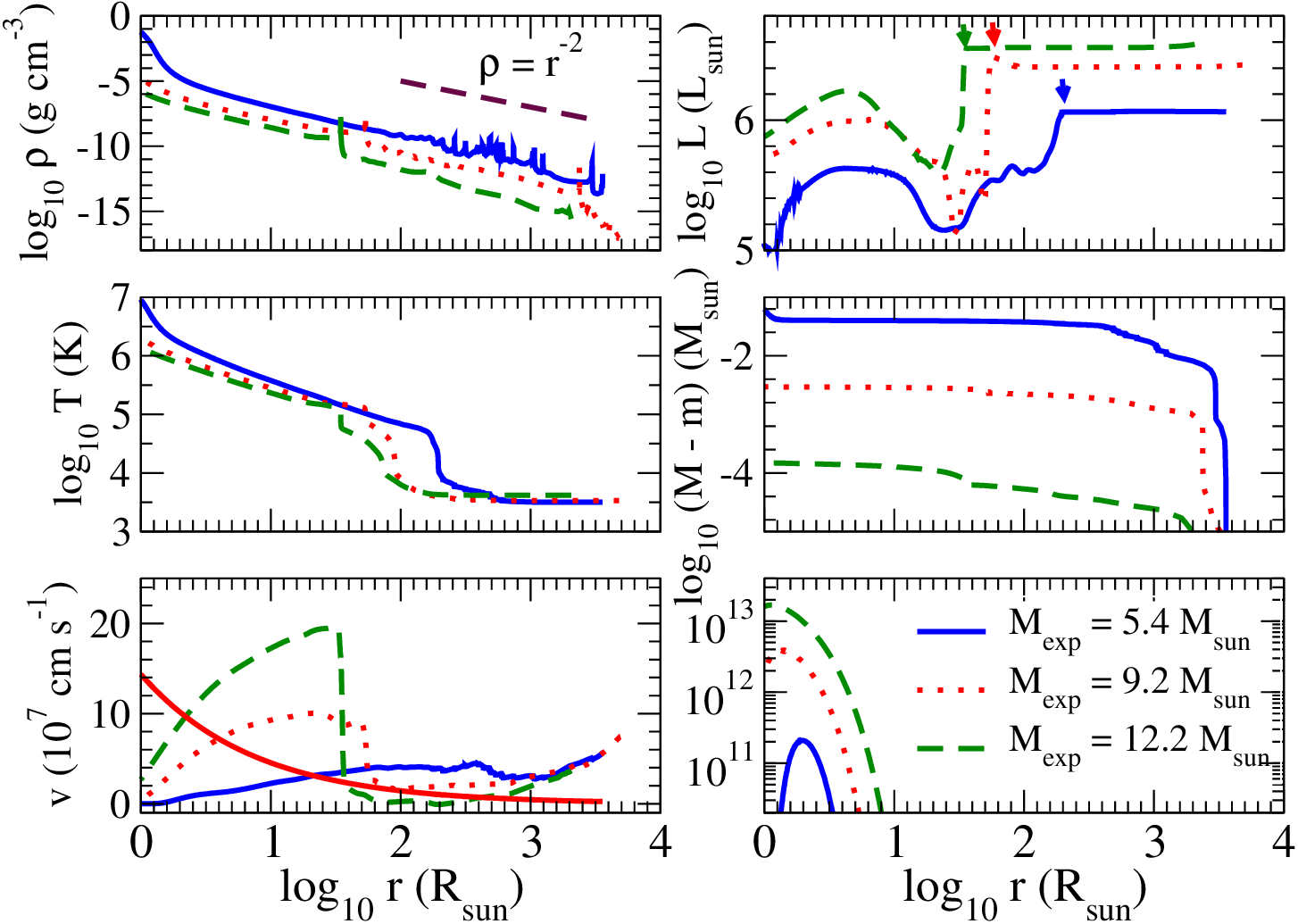}
\caption{
Similar to Figure \ref{fig:profile_tdep} but for models with a pre-explosion mass of $M_{\rm exp} = $ 5.44 (blue solid line), 9.2 (red dotted line) and 12.22 $M_{\odot}$ (green dashed line) respectively. An energy of $6 \times 10^{47}$ erg over a duration of 0.15 year is deposited during the simulation. In the top right panel, we add the arrows to indicate the formal photosphere ($\tau = 2/3$) radius. In the bottom left plot, the escape velocity (solid red line) corresponds to that of the model with $M_{\rm exp} = 5.44~M_{\odot}$.}
\label{fig:profile_mass}
\end{figure*}

We proceed to examine how the wave heating process affects helium stars of different mass. A higher progenitor mass leads to a more compact core with a smaller radius during oxygen burning, which results in an envelope of higher binding energy. In Figure \ref{fig:profile_mass} we compare models with a pre-explosion mass of $M_{\rm exp} =$ 5.44, 9.20 and 12.22 $M_{\odot}$ (corresponding to ZAMS models of $M_{\rm ini} =$ 25, 40 and 60 $M_{\odot}$ respectively, see Table \ref{table:mass}). For ease of interpretation, we assume a fixed energy deposition of $6 \times 10^{47}$ erg for a duration of 0.15 year, similar to the O-burning duration of a $M_{\rm ini} =$ 40 $M_{\odot}$ star. 

In Figure \ref{fig:profile_mass}, we plot the hydrodynamical profiles of these models at the end of these simulations. Higher stellar mass leads to a much lower amount of mass being ejected, from more than $10^{-2} ~M_\odot$ in the the $M_{\rm exp} =$ 5.44 $M_{\odot}$ model, down to about $10^{-4}~M_\odot$ in the the $M_{\rm exp} =$ 12.22 $M_{\odot}$ model. The outermost layers of the CSM extend to a few thousand solar radii in each case. Again the density profile falls off slightly steeper than $r^{-2}$, and the temperature profile flattens at a temperature of $T \approx 6000 \, {\rm K}$ in the optically thin outer regions of the ejecta.
The luminosity of the models is about ten to a hundred times of the original luminosity and increases with $M_{\rm exp}$.

\section{Physical Models of SN 2018gep}
\label{sec:interpret}

As discussed in the introduction, pulsation-driven mass loss is a well studied mechanism for explaining a wide range of observed CSM and its shock interaction implied in supernova light curves. In this section, we compare the CSM properties produced by the wave-driven mass-loss models and PPISN models. We also study how the two models compare to the CSM constraints derived from radiative transfer modeling of SN 2018gep in this work (See Section \ref{sec:radtran} for details) and in \cite{Ho2019}. 

\begin{figure}
\centering
\includegraphics*[height=6cm]{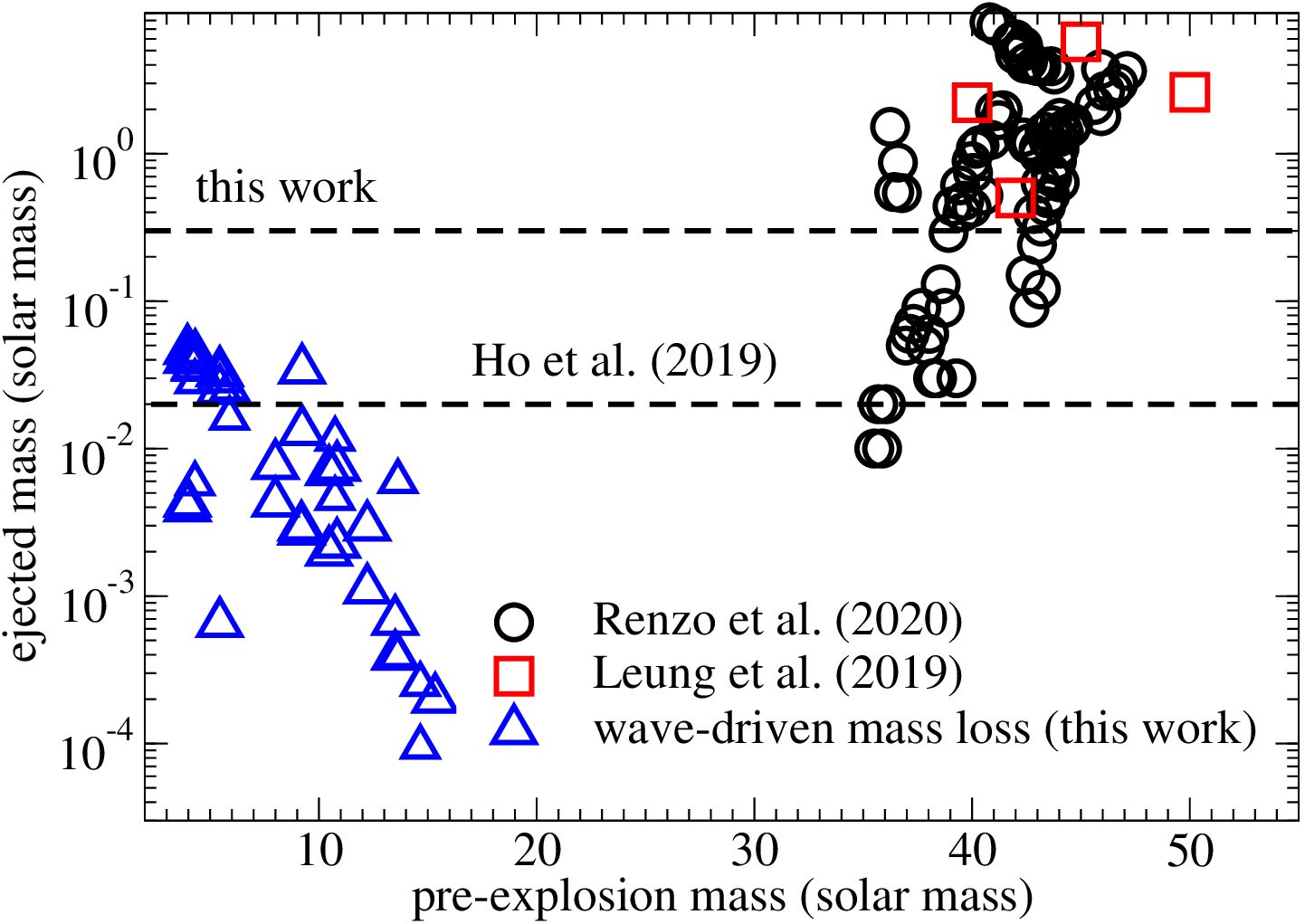}
\includegraphics*[height=6cm]{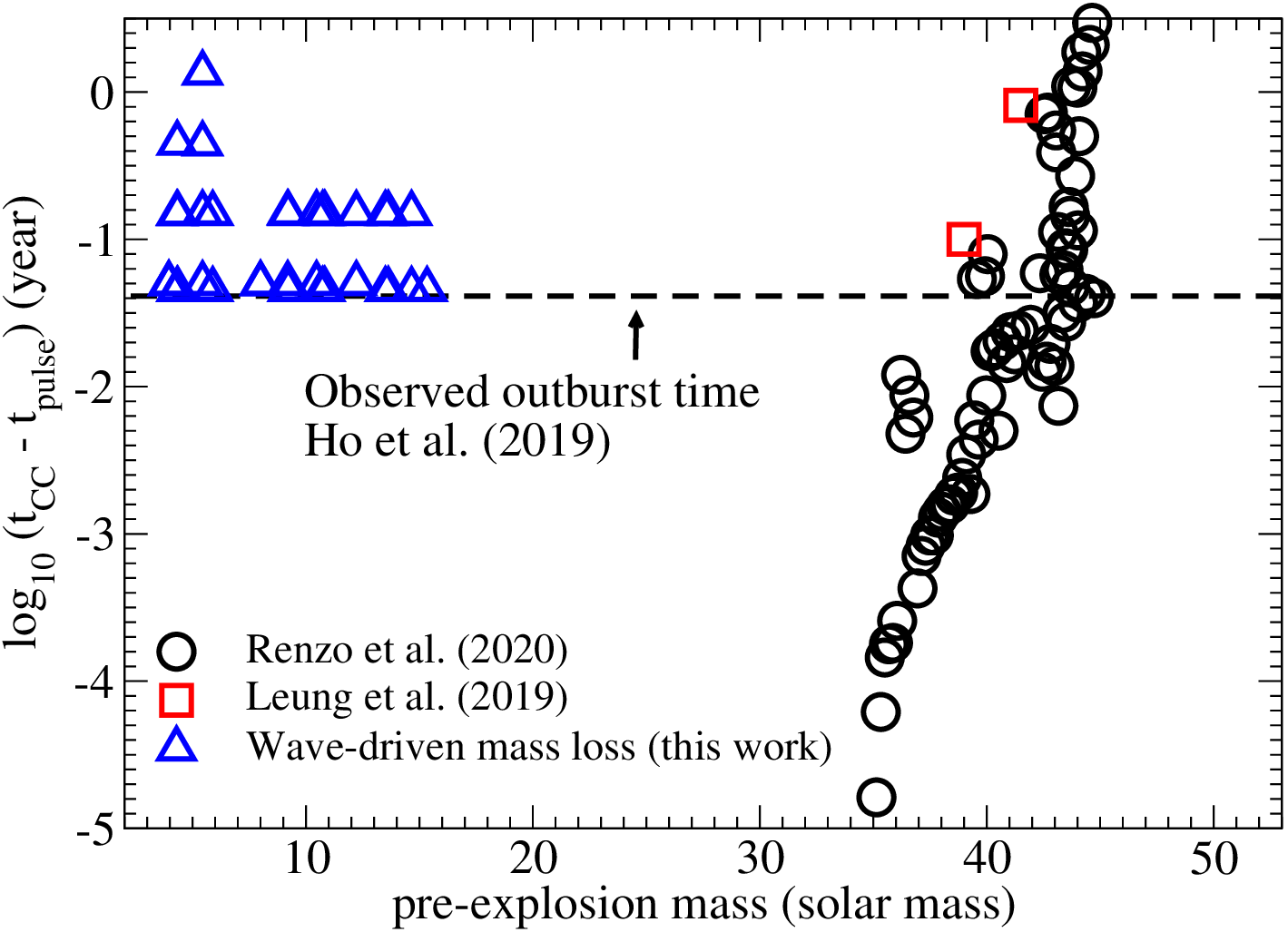}
\includegraphics*[height=6cm]{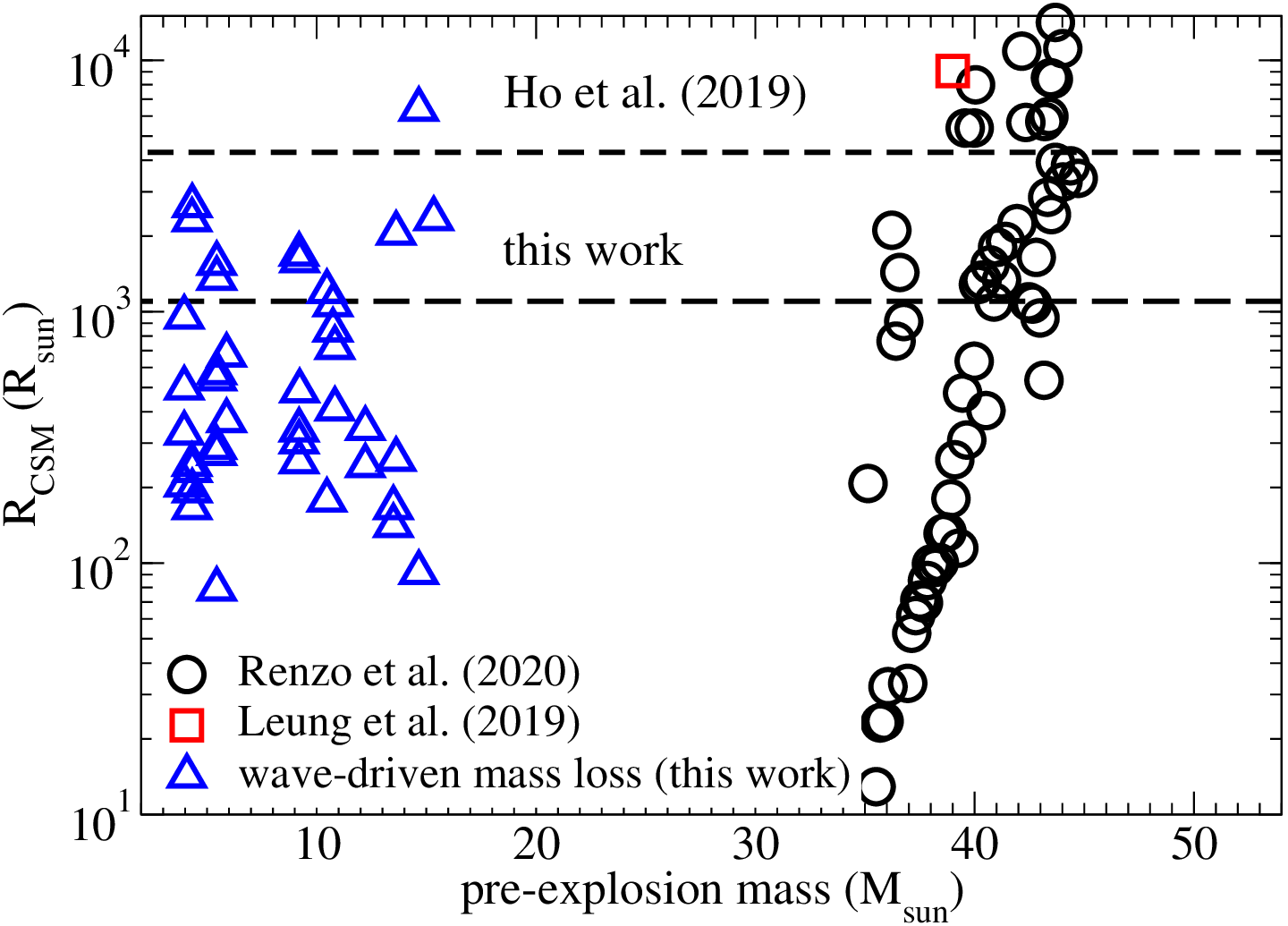}
\caption{
(top panel)
A comparison between the CSM properties in the wave-driven mass-loss regime and the pulsational pair-instability supernova regime. Horizontal lines correspond to values derived in \cite{Ho2019} and from our radiative transfer models in Section \ref{sec:radtran}, by using the observed bolometric light curve as a constraint. (middle panel) Same as the top panel but for the time between the mass ejection and core-collapse. (bottom panel) Same as the top panel but for the extracted mass-weighted CSM radius against pre-explosion mass. 
}
\label{fig:massloss_global}
\end{figure}

We compare the associated mass loss of the two models in Figure \ref{fig:massloss_global}, where we plot the total ejected mass against the pre-explosion mass of the star. Data for PPISN models come from prior results in the literature 
\citep{Renzo2020, Leung2019PPISN}. We can see two distinctive clusters in the figures. On the lower mass side (wave-driven mass loss), there is a clear trend in which the ejected mass decreases with pre-explosion mass, as discussed in the previous section. 
%which resemble with that in Figure \ref{fig:massloss}. 
For the PPISN models on the right hand side, we see a clear rising trend in which ejecta mass typically increases with pre-explosion mass. We also see the two sets of PPISN models mostly agree with each other. In both regimes, the trends are accompanied by scatter because the exact amount of mass loss is coupled to highly nonlinear or uncertain factors, including the temperature-sensitive carbon- and oxygen burning, or the amount of wave energy transport. For PPISN, uncertainties in the 1D treatment of dynamics and time-dependent convective energy transport further contribute to the observed scattering \citep{Renzo2020b}. 

The gap between the two clusters of models exists because we do not consider low metallicity models ($Z < 0.002$) for our wave-driven mass loss calculations, which use models with $Z=0.02$ or $Z=0.007$ so no progenitors with $M_{\rm exp} > 20~M_{\odot}$ form. \cite{Renzo2020b} uses a metallicity of $Z = 0.001$ for their PPISN models. The host galaxy of SN 2018gep has a metallicity of $Z \approx 0.004$. While the metallicity is important for determining the progenitor mass due to wind mass loss, metallicity is less important for wave-driven or pulsation-induced mass loss \citep{Renzo2020b}.
%The mass loss is in general insensitive to the initial metallicity of the star.}
While lower metallicity models can produce pre-explosion masses above this limit, we do not expect wave-driven mass loss to play an important role in CSM formation for those stellar models, based on the trend in this work.  Furthermore, until the pre-explosion mass approaches $35 \, M_\odot$, the models are not massive enough to trigger pair-creation instability, thus having no mass loss by pulsation.

We also plot two horizontal lines in the figure, corresponding to the CSM masses derived from the radiative transfer models of SN 2018gep in \cite{Ho2019} and in this work.
The wave-driven mass loss model cannot match the higher $M_{\rm CSM}$ value derived from our radiative transfer models, but it can match the value inferred from \cite{Ho2019} in progenitors with a pre-explosion mass $M_{\rm exp} =$ 4 -- 10 $M_{\odot}$. On the other side, both possible CSM masses can be explained by pulsation-induced mass-loss. The smaller CSM mass from  \cite{Ho2019} can be explained by PPISN models with $M_{\rm ini} \sim 35 \, M_\odot$; while our higher inferred CSM mass matches models with a pre-explosion mass $M_{\rm exp} \sim 40~M_{\odot}$.

In the middle panel of Figure \ref{fig:massloss_global}, we show the time delay between the mass ejection and the onset of core-collapse for the same sets of models. For our wave-driven mass loss models, the energy deposition time is a model parameter as discussed in the previous section.
For pulsation driven mass loss, the time delay is defined as the time between the first pulse and core-collapse. The PPISNe have a wide range of time delays which span from $\sim 10^{-5}  - 10^{1}$ year.
A time delay in which the collapse occurs before the pulse arrives at the surface is defined as zero.
We also plot the observed time delay inferred from pre-explosion imaging of the progenitor of SN 2018gep in \cite{Ho2019}, which revealed an outburst roughly 15 days (0.04 years) before the final explosion. Many of the PPISN models in \cite{Renzo2020} are near the observed time delay of $\sim \! 0.04 \, {\rm yr}$ for SN 2018gep, but this time delay is also similar to that expected for many wave-driven mass loss models. We note that, despite the very different mass ejection physics in the two classes of models, they are both primarily powered by energy from O-burning, which also explains the similarity of the range of time delay.

In the bottom panel of Figure \ref{fig:massloss_global}, we plot the characteristic CSM radius for our models and the PPISN models of \cite{Renzo2020}. We define the characteristic CSM radius as the mass-averaged radius of material which has a positive energy.
For the PPISN models, we estimate $R_{\rm CSM} = <v> \Delta t$, where $\Delta t$ is the same time delay used in the middle panel, and $<v>$ is the mass-averaged ejecta velocity from \cite{Renzo2020}. 

The wave-driven models do not show an obvious trend in the CSM radius, though it is possible that more realistic models (which self-consistently calculate wave energy transport as a function of time) would exhibit a trend. In any case, they span a range in CSM radius from roughly $10^2 - 10^4 \, R_\odot$. This range includes the estimated CSM radius from radiative transfer modeling of $R_{\rm CSM} \sim 10^3 \, R_\odot$ in this work (again see Section \ref{sec:radtran}), and $R_{\rm CSM} \sim 4000 \, R_\odot$ from \cite{Ho2019}.
On the other hand, the PPISN models predict a wide range of CSM radii from roughly $10 - 10^5 ~R_{\odot}$. They also cover both the CSM radius suggested in this work and in \cite{Ho2019}.

By combining all three plots, we determine the most likely progenitor model of SN 2018gep. The wave-driven models can match the ejecta mass and time delay from \cite{Ho2019}, but they struggle to match the larger $R_{\rm CSM}$ from that work. Additionally, the wave-driven mass loss models cannot match the larger $M_{\rm CSM}$ inferred from our radiative transfer models in Section \ref{sec:radtran}. 

Most PPISN models exceed the CSM mass found from \cite{Ho2019}, but they can match the larger CSM mass from our radiative transfer models.
The time delay can also be comparable with that observed. However, the PPISN models which match the CSM mass from this work typically exceed the CSM radius from this work by a factor of a few. Those that match the CSM mass from \cite{Ho2019} have smaller CSM radii than that inferred from \cite{Ho2019}, and their outburst times are shorter than that observed. In conclusion, there are no individual PPISN models which perfectly match the CSM mass, radius, and pre-explosion time, but there are several at masses of $M_{\rm exp} = 37-44 \, M_\odot$ that are within a factor of a few of the inferred values.

\cite{Ho2019} also found an ejecta mass of $M_{\rm ej} \sim 8~M_{\odot}$ from modeling the late-time light curve, implying a pre-explosion mass near $10 \, M_\odot$ if a neutron star is formed during the explosion.
Such progenitors struggle to eject enough CSM mass by the wave-driven mass loss mechanism,
unless we consider models with an optimistic amount of deposited energy ($\gtrsim \! 10^{48}$ erg). However, the inferred ejecta mass is much lower than the pre-exposion mass required for pulsation-induced mass loss. Hence, PPISN models can only match the ejecta mass if a large fraction of the progenitor mass collapses into a black hole, or if the SN ejecta is not spherically symmetric (e.g., most of the SN light arises from high-velocity bipolar lobes containing a small fraction of the ejecta mass).

\section{Light Curve Fitting of SN 2018gep}
\label{sec:radtran}

\subsection{Light Curve Modeling}
In this section, we use radiative transfer modeling to derive constraints for the CSM based on the observational data of SN 2018gep.
We use the supernova explosion code \citep[SNEC (Supernova Explosion Code)]{Morozova2015}
to compute the post-explosion light curve. This code follows the prototype presented in \cite{Bersten2011,Bersten2013}.
The code has been widely used in the literature so we 
refer the reader to the instrumentation paper about the detailed implementation and its test cases. In this section, we choose $M_{\rm ini} = 25~M_{\odot}$ $(M_{\rm exp} = 5.44~M_{\odot})$ as our default pre-explosion model. We also included a model of $M_{\rm ini} = 50~M_{\odot}$, whose He-core of $M_{\rm He} =$ 17.45 $M_{\odot}$ is reduced to a pre-explosion mass of $M_{\rm exp} = $ 10.46 $M_{\odot}$ by winds (see Table \ref{table:mass}). The model is used for comparison in Figure \ref{fig:LC_compare_Ho}. The ejecta mass is chosen by selecting the mass cut. Here we only briefly review the our input.

After constructing ordinary stellar evolutionary models (i.e., models without any wave heating), we extract the density and chemical composition profile from MESA. Then we determine some input parameters necessary for SNEC. This includes,
(1) the inner mass cut $M_{\rm cut}$, 
(2) the explosion energy $E_{\rm exp}$
(3) the $^{56}$Ni distribution inside the star $X(^{56}$Ni). 
%and the inner mass cut. 
The minimum mass cut is the mass coordinate of the Fe-core, assuming the Fe-core to be promptly collapsing due to gravitational instability. A larger mass-cut corresponds to an explosion with fallback or an aspherical explosion.

\begin{figure}
\centering
\includegraphics*[height=6cm, height=6cm]{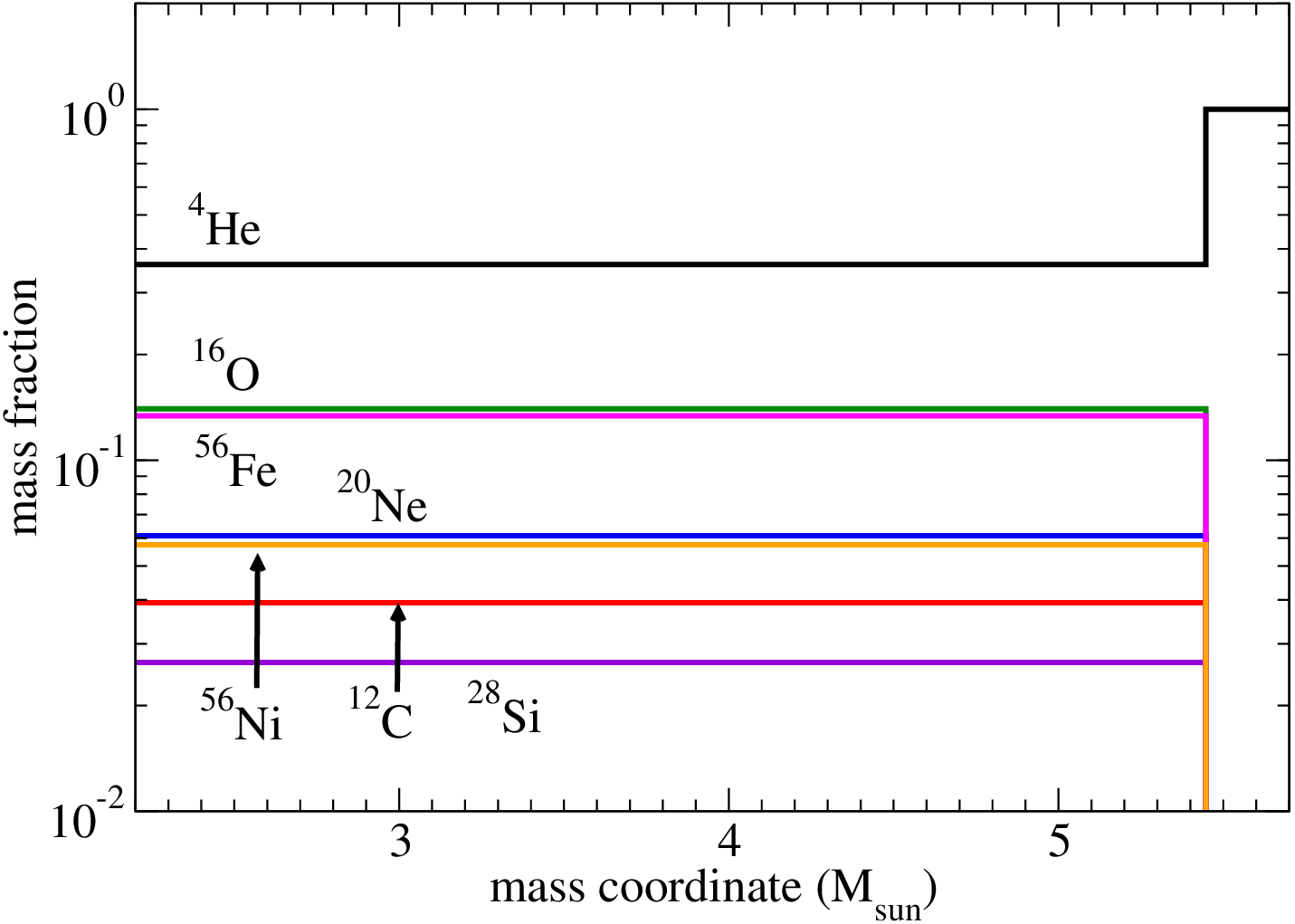}
\caption{The chemical abundance profile of representative elements used for the default model. Matter at $M(r) > 5.44$ belongs to the CSM. }
\label{fig:init_abund}
\end{figure}

For ease of model comparison, we do not use the exact progenitor models from Section \ref{sec:Progenitor}. Instead, we add parameterized CSM to our models, with properties motivated by wave-driven and pulsation driven mass loss. Outside the original stellar surface, we add a CSM that extends to radius $R_{\rm CSM}$ with mass $M_{\rm CSM}$. The CSM is initially isothermal with a density dependence $\rho \propto 1/r^2$. For PPISN models, the extended and smooth mass ejection history reported in \cite{Leung2019PPISN} (Figs. 23 -- 26) also suggests a similar CSM profile. These parameters should not be over-interpreted since the real CSM density profile is not well predicted by stellar evolution models.
Similar to \cite{Leung2020COW}, we assume the ejecta is uniformly mixed. In Figure \ref{fig:init_abund} we plot the chemical abundance of the default model for illustration.

We emphasize that the light curve models are most sensitive to the ejecta-CSM interaction and are not very sensitive to the progenitor model. Therefore, for a given ejecta mass and composition, the results and constraints on CSM are applicable to both wave-driven mass loss and PPISN models. In Section \ref{sec:aspherical} we further discuss how the aspherical explosion of massive stars is connected to these properties.

\subsection{Best-fit Model}

\begin{figure*}
\centering
\includegraphics*[height=6cm, height=6cm]{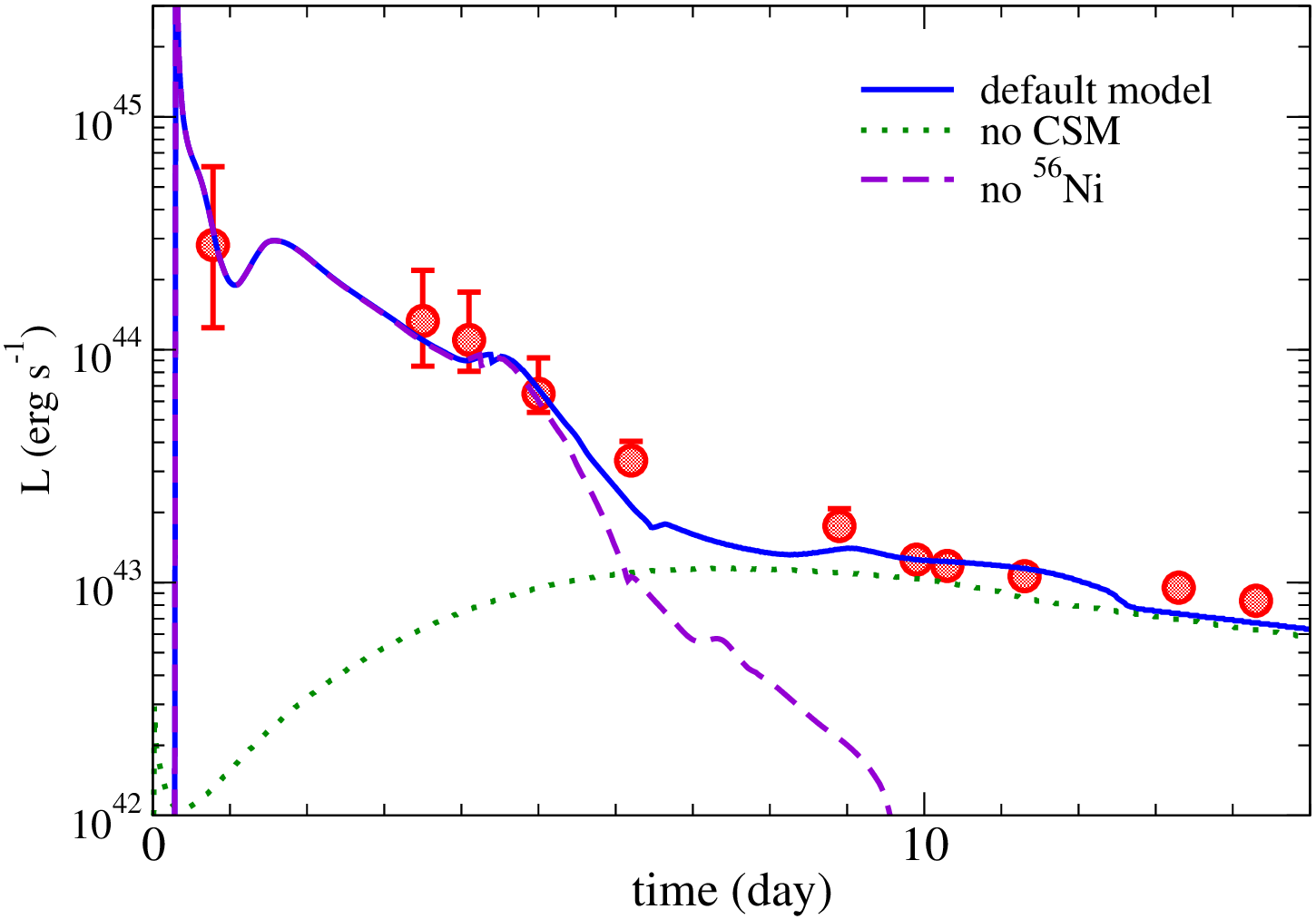}
\includegraphics*[height=6cm, height=6cm]{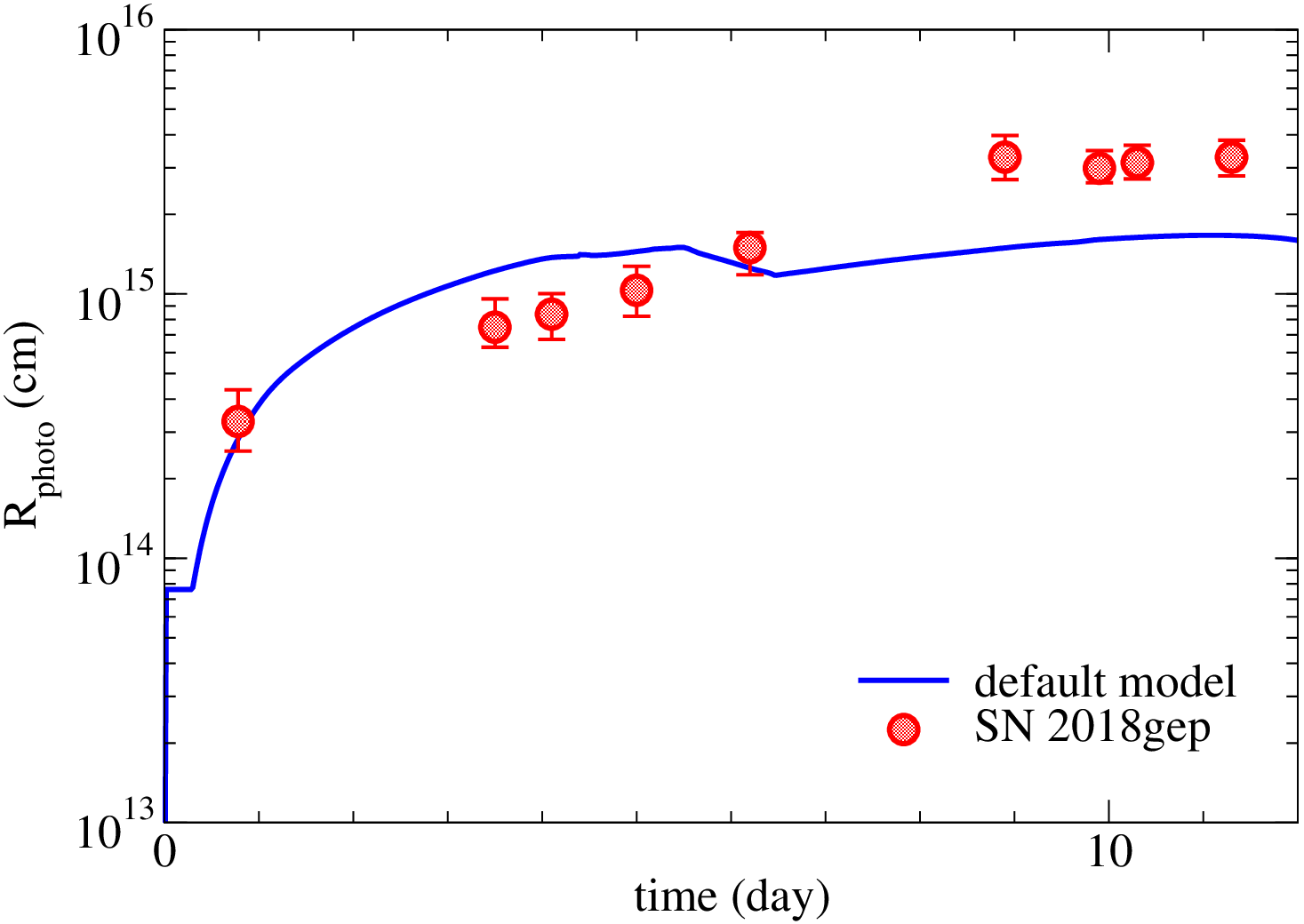}
\includegraphics*[height=6cm, height=6cm]{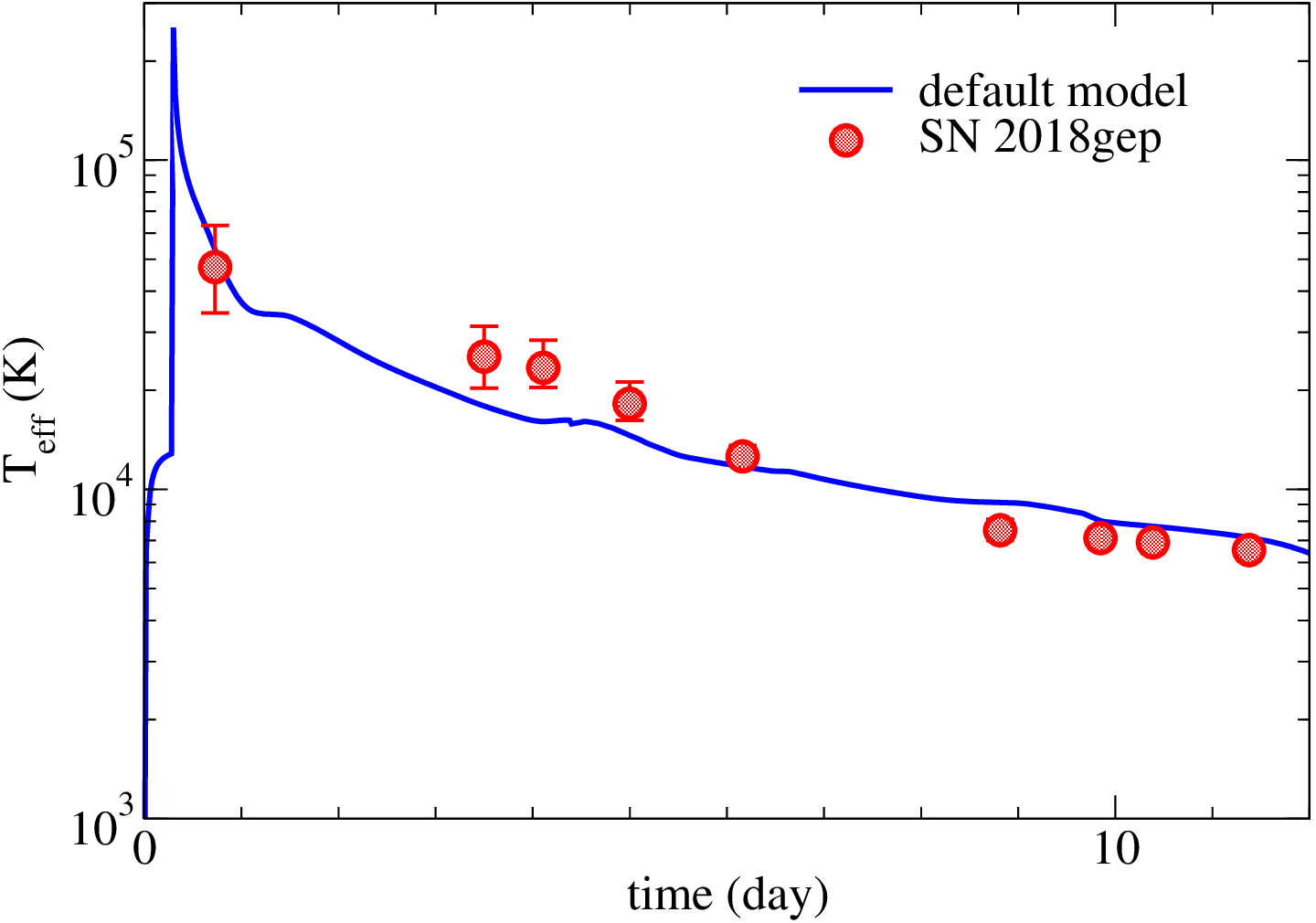}
\includegraphics*[height=6cm, height=6cm]{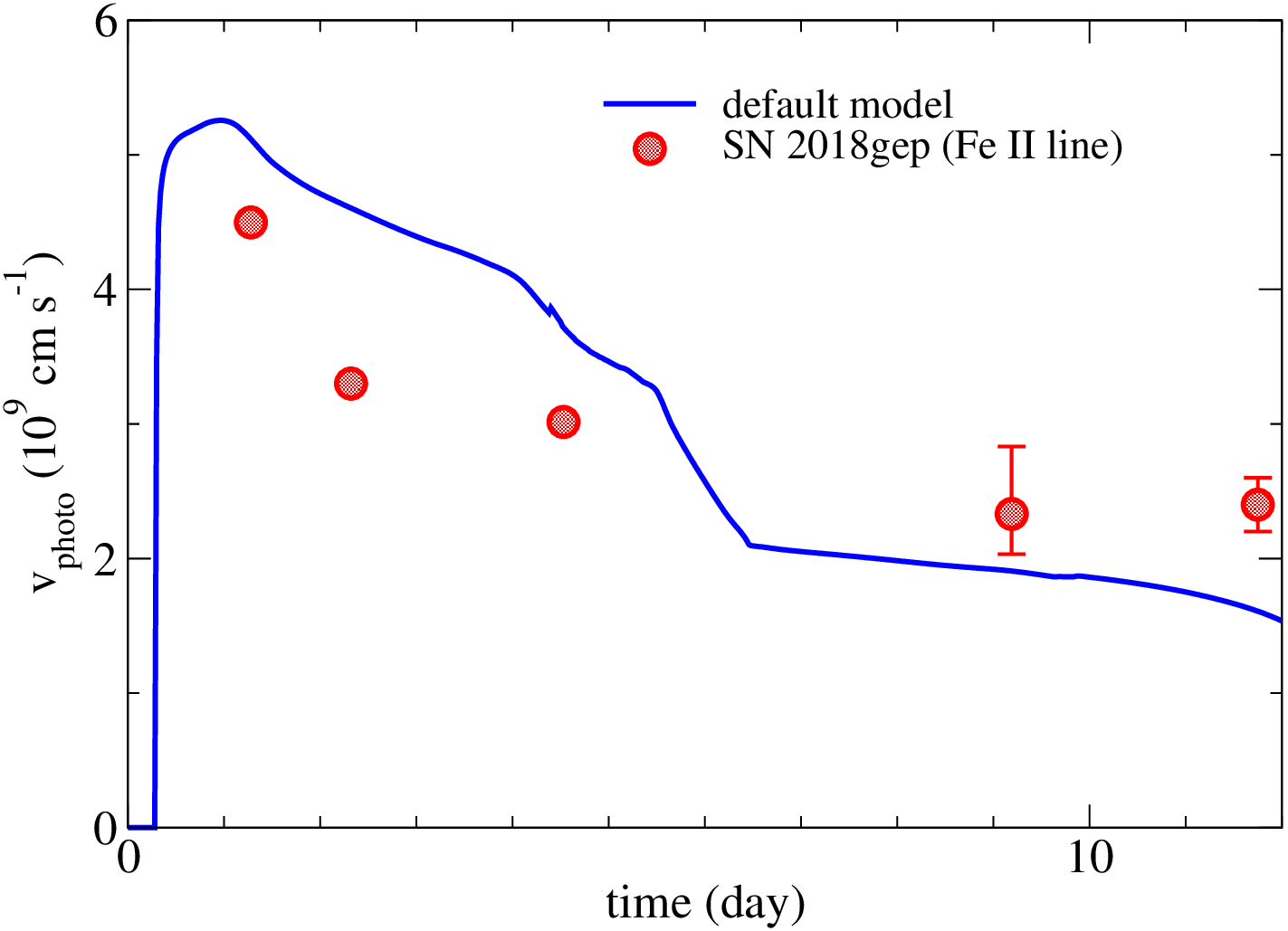}
\caption{
(top left) The light curve of the default model (blue solid line), a contrasting model with only CSM (purple dashed line) or only $^{56}$Ni (green dotted line). The red circles correspond to the observed values from SN 2018gep \citep{Ho2019}. (top right) The photosphere radius of the default model. (bottom left) The effective temperature of the default model. (bottom right) The photosphere velocity of the default model.
}
\label{fig:benchmark_model}
\end{figure*}

First we present our default model, which fits many of the features in the bolometric light curve of SN 2018gep. To find a good match, we need to search over the progenitor mass $M_{\rm ini}$, ejecta mass $M_{\rm ej}$, CSM mass $M_{\rm CSM}$, CSM radius $R_{\rm CSM}$, explosion energy $E_{\rm exp}$, and $^{56}$Ni mass in the ejecta $M(^{56}$Ni). In such a high dimensional parameter space, we do not attempt to fit the light curve perfectly. Instead, we try to understand how the light curve shape depends on each of the parameters.
%so that future extension of this work can apply the finding of this work as a general direction.
Table \ref{table:default} lists the parameters of our default model. 

\begin{table}
\begin{center}
\caption{The parameters of our default progenitor and explosion model for SN 2018gep.}
\begin{tabular}{c|c}
\hline
parameter & value \\ \hline
pre-explosion mass ($M_{\rm exp}$) & 5.44 $M_{\odot}$ \\
$M_{\rm He}$, $M_{\rm C}$ & 6.56, 3.00 $M_{\odot}$ \\
%metallicity (Z) & 0.02 $Z_{\odot}$ \\
explosion energy ($E_{\rm exp}$) & $1 \times 10^{52}$ erg \\
Ejecta mass ($M_{\rm ej}$) & 2.0 $M_{\odot}$ \\
CSM mass ($M_{\rm CSM}$) & 0.3 $M_{\odot}$ \\
CSM radius ($R_{\rm CSM}$) & 1100 $R_{\odot}$ \\
Nickel mass ($M(^{56}$Ni)) & 0.33 $M_{\odot}$ \\ 
CSM slope & -2 \\ \hline

\end{tabular}
\label{table:default}
\end{center}
\end{table}

In Figure \ref{fig:benchmark_model}, we plot the bolometric light curve (top left), photospheric radius (top right), effective temperature (bottom left) and photospheric velocity (bottom right) of our default model. For the bolometric light curves, we also include control experiments to demonstrate how the $^{56}$Ni-decay (green dotted line) and the pure CSM interaction (purple dashed line) contribute individually to the formation of the full light curve (blue solid line). SN 2018gep shows a two-component structure in the light curve. In the first several days, there is a plateau in the light curve where the luminosity is $\sim \! 10^{44}$ erg s$^{-1}$. During this time, the luminosity is almost entirely created by thermal diffusion out of the shock-heated CSM. Around day 5, the CSM contribution decreases rapidly as the CSM becomes optically thin. The light curve falls more steeply until it flattens again when $^{56}$Ni-decay dominates the luminosity.

In this work, we define $t = 0$ as the moment of explosion of the model, which takes place at UT 2018 September 8 20:20. This is about 0.3 days before the definition of $t=0$ in \cite{Ho2019} based on the g-band light curve brightening. Indeed, shock breakout occurs roughly 0.3 days after explosion in our default model, roughly consistent with the observed time of rapid brightening.

The observed photospheric radius $R_{\rm ph}$ of SN 2018gep shows a steady rise from an initially high value of $\sim 3 \times 10^{14}$ to $\sim 3 \times 10^{15}$ cm. Our models show similar evolution, albeit with a more rapid initial increase in $R_{\rm ph}$ and a less rapid late-time increase in $R_{\rm ph}$. After day 5, there is a small but sudden drop in $R_{\rm ph}$ in our models when the shock-heated CSM becomes transparent.
%The exact photosphere radius then is smaller than the observed one. 
For the effective temperature $T_{\rm eff}$, the default model follows the trend of SN 2018gep fairly well. The model drops from $\sim 10^5$ K on day 1 to $\sim 10^4$ K at day 6, falling more gradually at later times. The largest discrepancy appears to be at the points between day 3 -- 6, where our model under-predicts the temperature and over-predicts the photospheric radius. The imperfect fit could also indicate a CSM with slightly larger radial extent or different density structure could better match the data. 

We note the small bumps in the light curve around day 1.5 and day 4.5. These bumps correspond to short temperature plateaus at the photosphere which occur at the first and second recombination of helium. The nearly constant photospheric temperature but increasing photospheric radius creates small and brief increases in the bolometric luminosity.

Lastly, we examine the evolution of the photospheric velocity, which is defined in the models as the Lagrangian velocity at the photosphere. The observed velocities of \cite{Ho2019} are measured using the C/O lines at early times (which are difficult to measure) and the Fe-II line at late times, which typically forms at smaller optical depth (and hence larger velocity) than the blackbody photosphere \citep{Morozova2020}. At most times, the model velocities are within $\sim$20\% of those observed. The model photosphere expands at a very high velocity of $\sim 3 - 4 \times 10^9$ cm s$^{-1}$ at early times when it is located within the CSM, but it falls sharply to about $2 \times 10^9$ cm s$^{-1}$ when the CSM becomes optically thin.  This sharp drop is not apparent in the data, though the sparseness and large uncertainties of the data prevents a detailed comparison.

\subsection{Hydrodynamical Evolution}

\begin{figure*}
\centering
\includegraphics*[width=8cm, height=6cm]{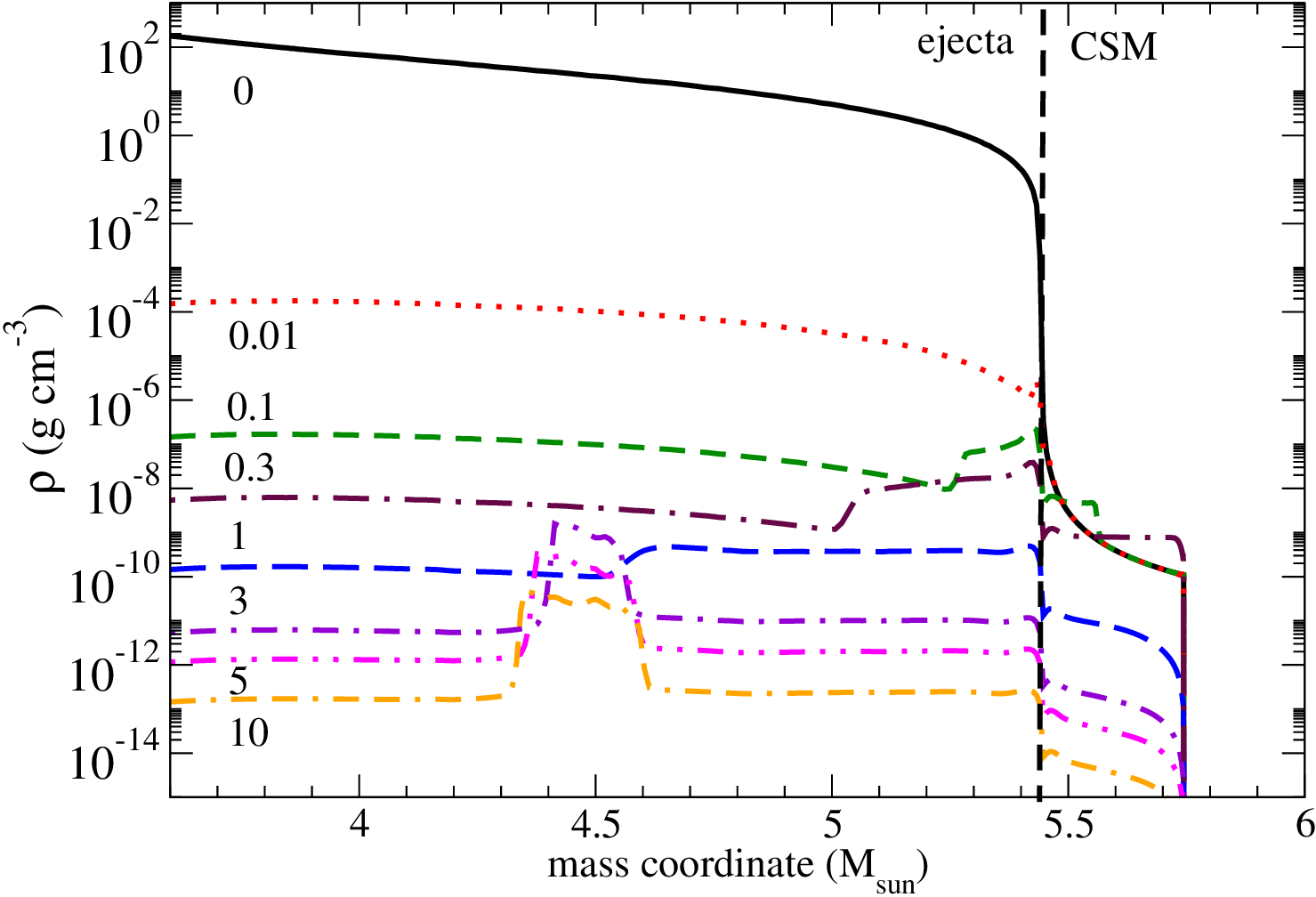}
\includegraphics*[width=8cm, height=6cm]{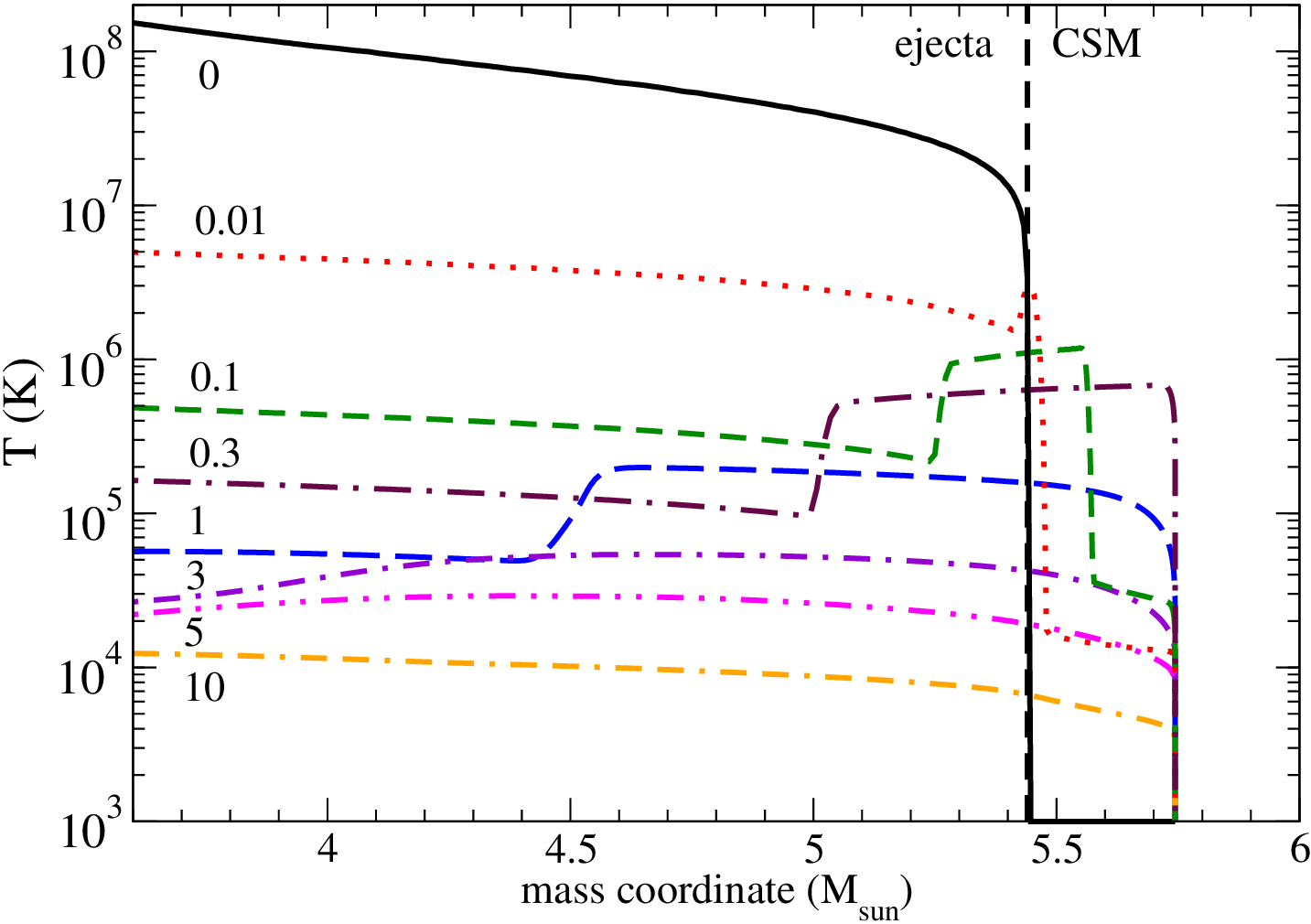}
\includegraphics*[width=8cm, height=6cm]{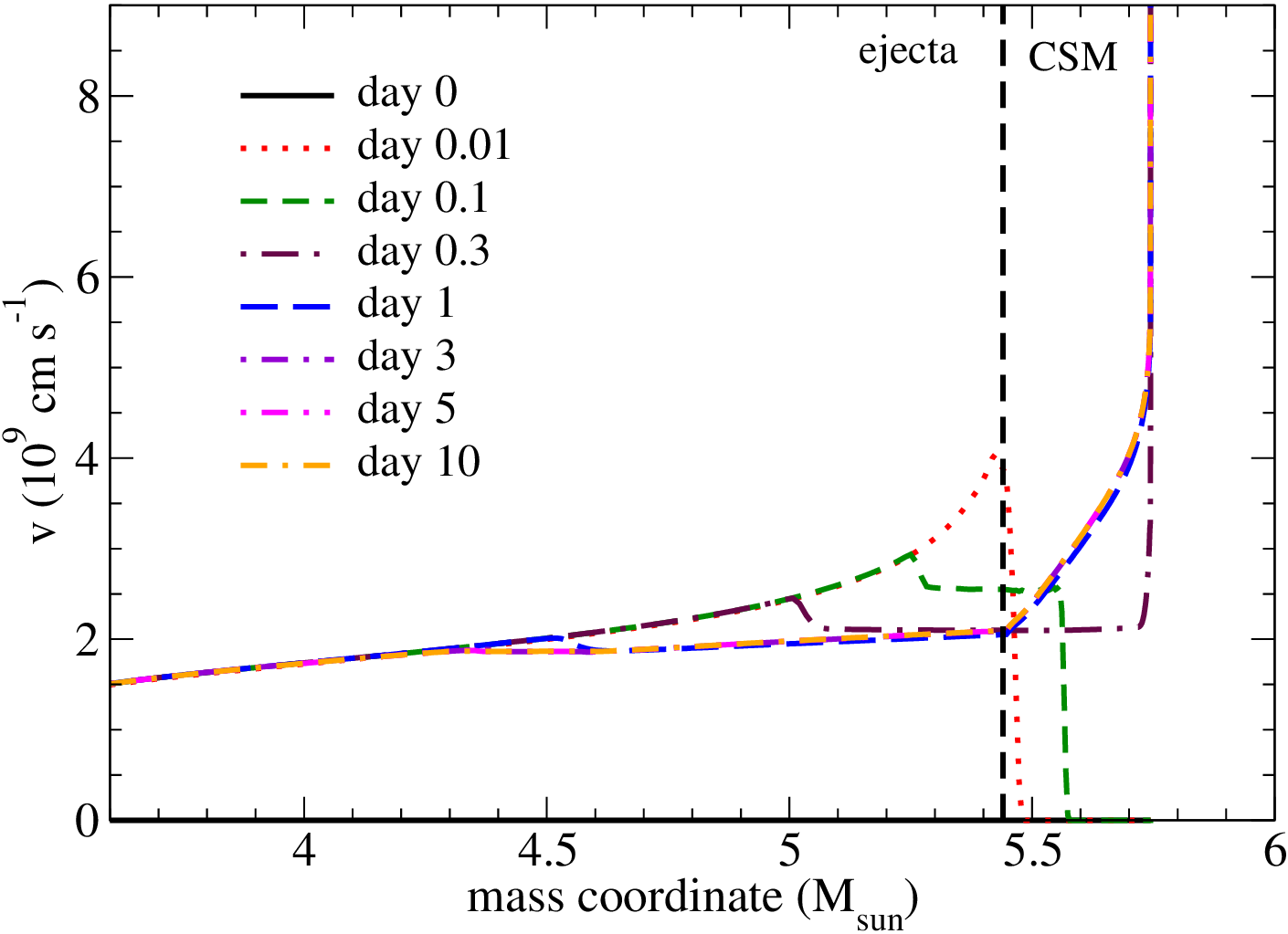}
\includegraphics*[width=8cm, height=6cm]{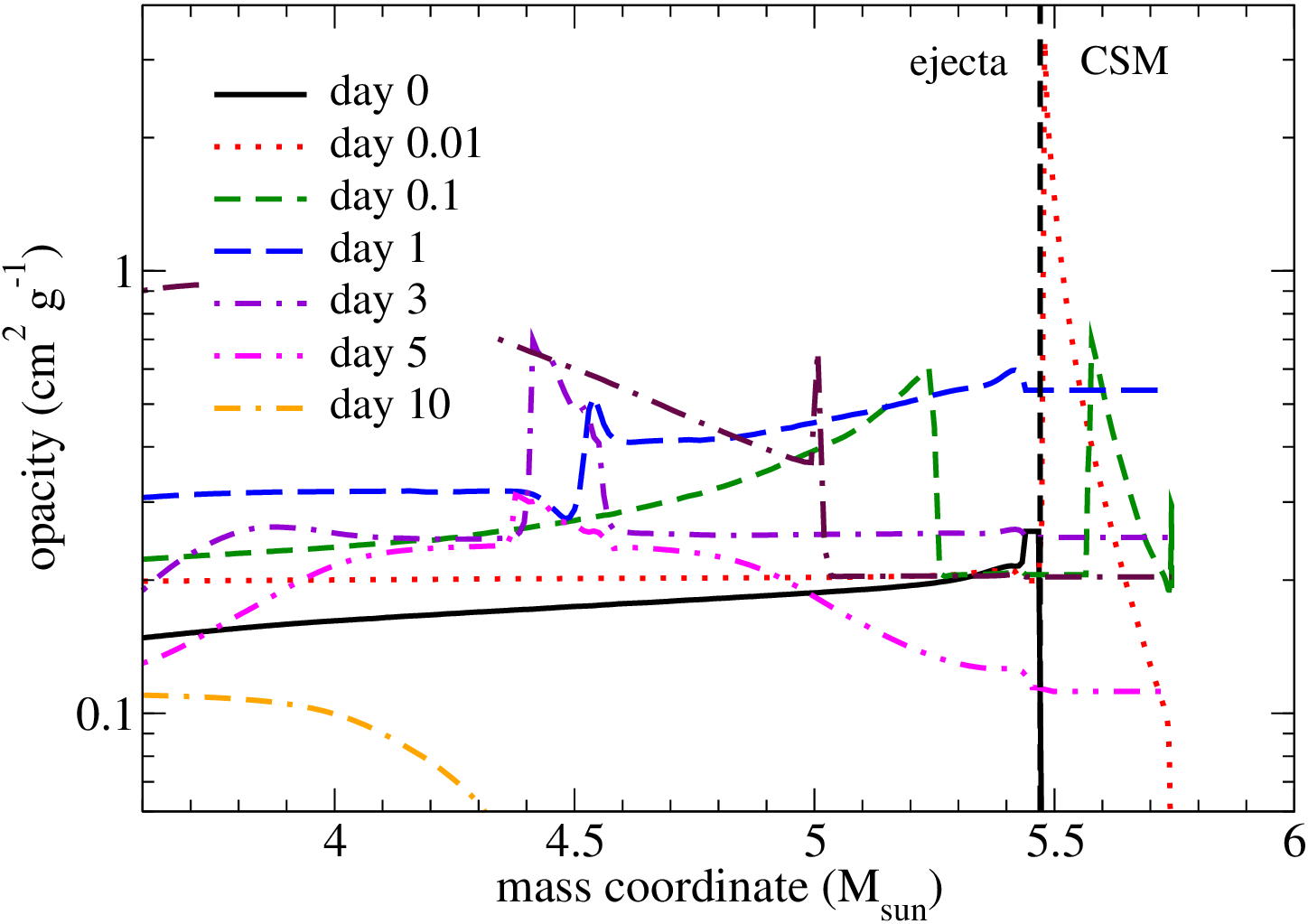}
\includegraphics*[width=8cm, height=6cm]{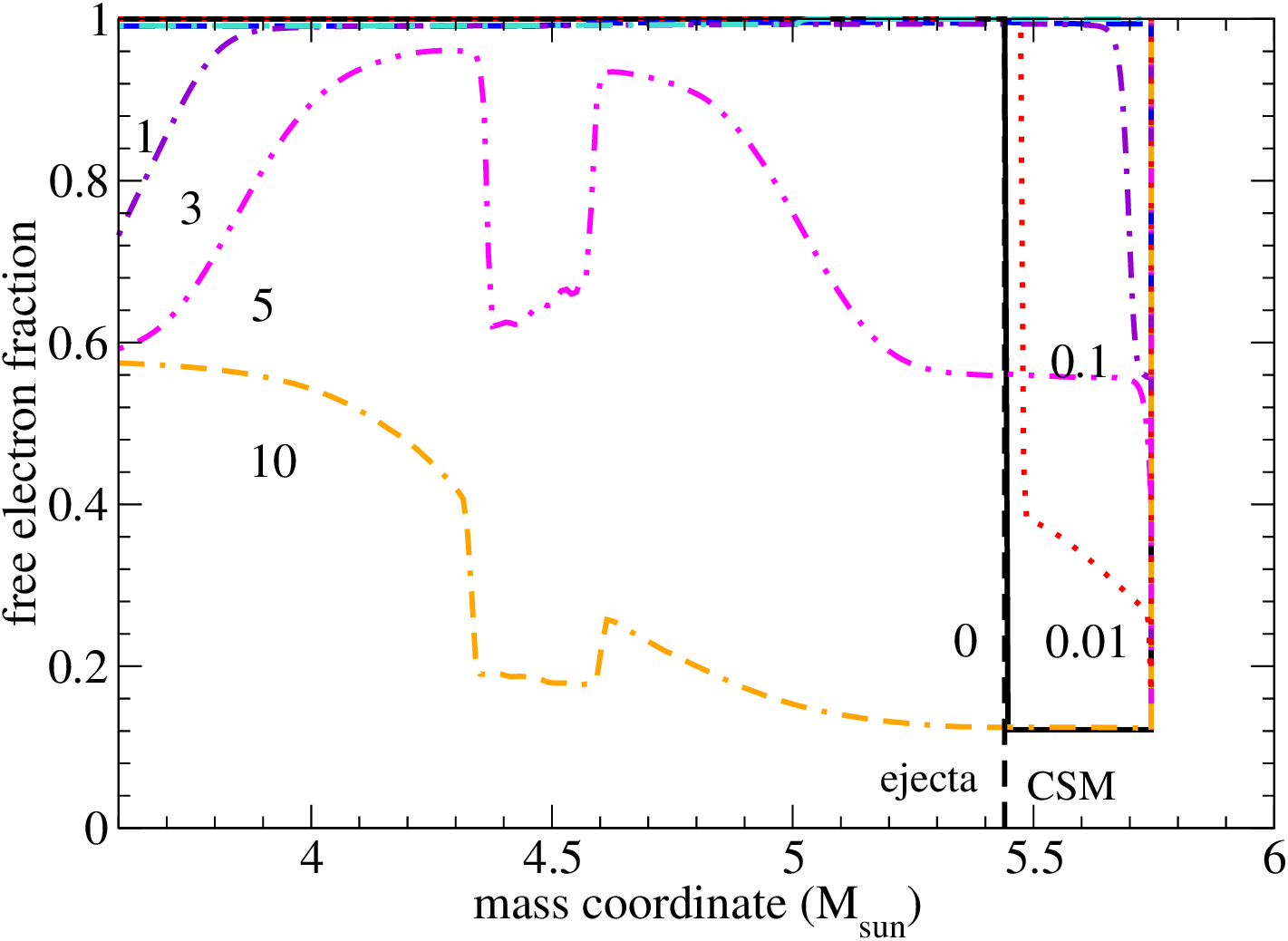}
\includegraphics*[width=8cm, height=6cm]{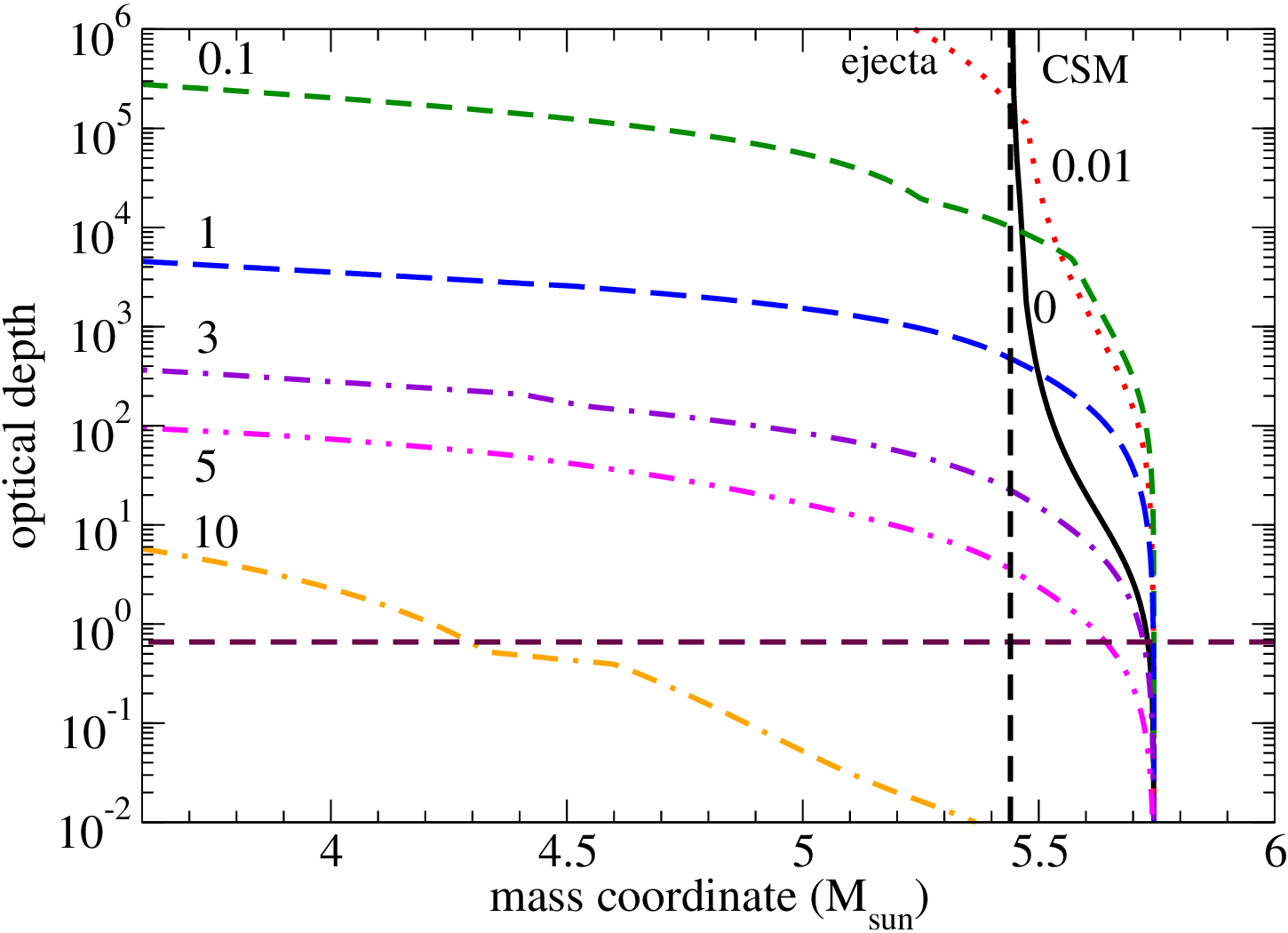}
\caption{
Hydrodynamical evolution of the default model including 
density (top left), temperature (top right), velocity (middle left), opacity (middle right), free electron fraction (bottom left), and optical depth (bottom right) at day 0 (black solid), 0.01 (red dotted), 0.1 (green dashed), 1 (blue long-dashed), 3 (purple dot-dashed), 10 (magenta dot-dot-dashed) and 15 (orange dot-dash-dashed) respectively. In the optical depth plot, we also add a horizontal line $\tau = 2/3$ whose intersection with the profiles defines the location of the photosphere. 
The vertical dashed line in all plots denotes the boundary between the SN ejecta and the CSM.}
\label{fig:benchmark_hydro}
\end{figure*}

We now study how the CSM shock cooling takes place in the default
model by examining the evolving hydrodynamical structure. Figure \ref{fig:benchmark_hydro} shows the density (top left), temperature (top right), velocity (middle left), the opacity (middle right), the free electron fraction (bottom left), and optical depth (bottom right) at several times after the explosion. Note that the innermost ejecta layers have a mass coordinate of $3.4 \, M_\odot$ due to the mass cut used in this model. The initial density profile is that of the stellar model and the constant density CSM.
When the shock hits the CSM, it creates a density bump at the contact discontinuity, which gradually vanishes as the star expands.  
Afterward, the density decreases steadily due to nearly homologous expansion, apart from another density bump that develops near the middle of the ejecta after day 3 at $M(r) \sim 4.3~M_{\odot}$. This is produced by a reverse shock propagating inwards from the contact discontinuity with the CSM. The compression associated with the reverse shock creates a density bump. The reverse shock evidently dies out at $M(r) \sim 4.3~M_{\odot}$, but the density bump is sustained long afterward.

As expected, the velocity increases with mass coordinate, and the star reaches locally homologous expansion by day 1. Before that, at day 0.01 and 0.1, we capture the moment when the shock hits the CSM and propagates through it. The majority of matter in the progenitor star (up to $5.44~M_{\odot}$) expands with a velocity $\sim 2 \times 10^9$ cm s$^{-1}$, while the low-density CSM obtains much larger velocities of $3 - 4 \times 10^{9}$ cm s$^{-1}$. The extremely thin layer of outer CSM can reach as high as $10^{10}$ cm s$^{-1}$.

Like the density, temperature generally decreases with time due to (initially adiabatic) expansion. The early shock-CSM collision leads to a hot outer layer of $T > 10^5$ K at early times. A temperature bump propagates inward due to the reverse shock propagating through the inner ejecta, which weakens as the star expands. At late times, the ejecta approaches isothermality as it becomes optically thin.

The opacity profile is closely linked to the temperature profile.
After shock heating, the opacity rises by a factor of 2 -- 3 where the shock-heated CSM is the most opaque. The opacity is typically largest where the temperature is $\sim \! 10^5 \, {\rm K}$ and free-free absorption dominates. At much larger temperatures, electron scattering dominates the opacity and $\kappa \approx 0.2 \, {\rm cm}^2/{\rm g}$ for the hydrogen-free ejecta. At much smaller temperatures ($T \lesssim 10^4 \, {\rm K}$), the ejecta recombines and the opacity plummets. This occurs starting around day 5, after which the free electron fraction and opacity decrease sharply. 

Finally we examine the optical depth evolution in the bottom right panel of Figure \ref{fig:benchmark_hydro}, where the horizontal line at $\tau = 2/3$ defines the photosphere. In the first 3 days, the photosphere lies in the outermost layers because the shocked heated CSM is opaque. After day 3, the shocked CSM cools and begins to recombine. This causes the optical depth to decrease rapidly such that the photosphere recedes, consistent with the sudden drop of photospheric radius on day 7 in Figure \ref{fig:benchmark_model}. By day 10, the photosphere has receded into the stellar ejecta below the CSM, and by day 15 nearly all of the ejecta is optically thin.

\subsection{Parameter-Dependence of Light Curve Models}
 
In this section, we examine how parameter variations away from 
the default model affect the shape of the light curve.

\subsubsection{Dependence on Explosion Energy}

\begin{figure}
\centering
\includegraphics*[ height=6cm]{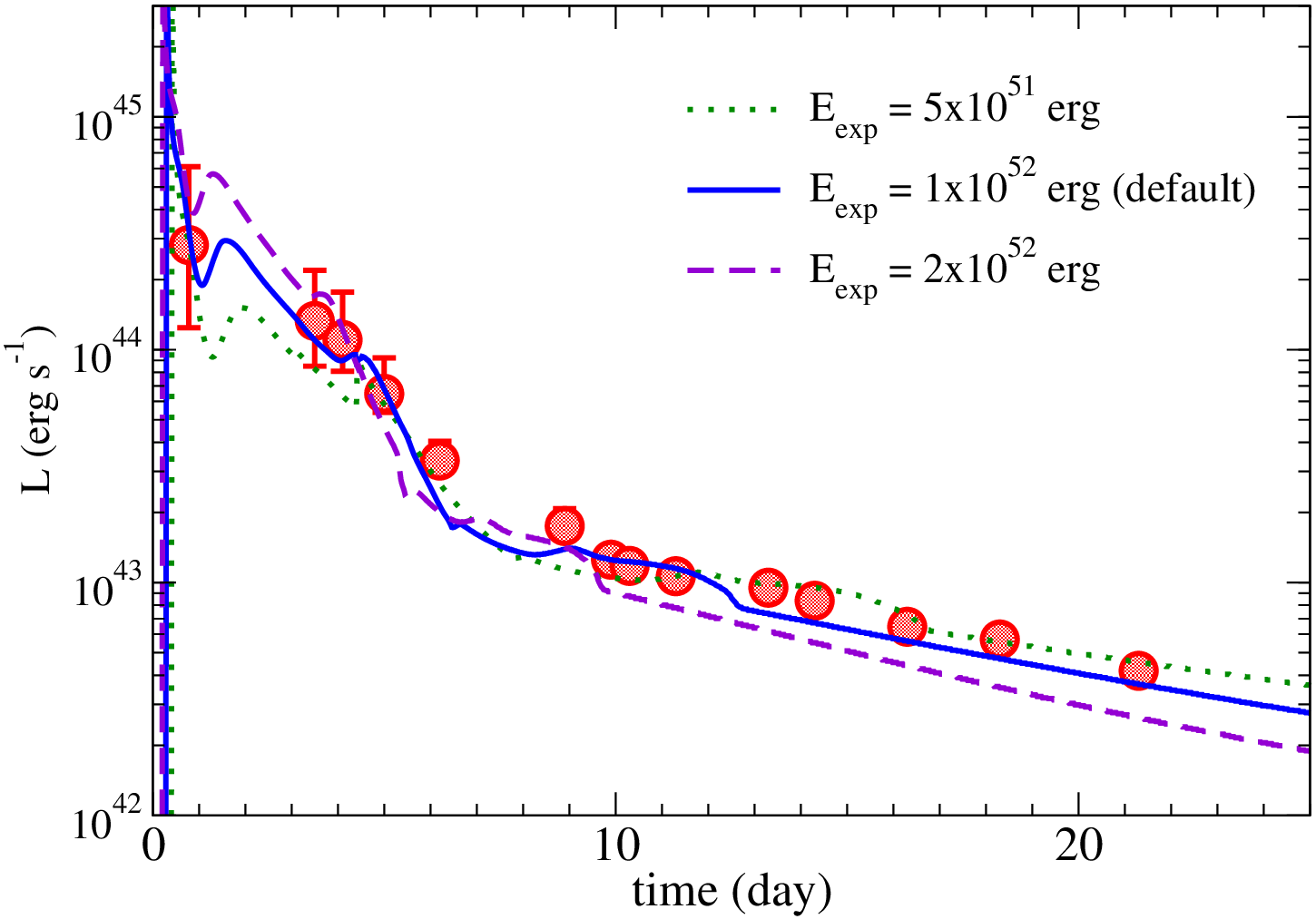}
\includegraphics*[ height=6cm]{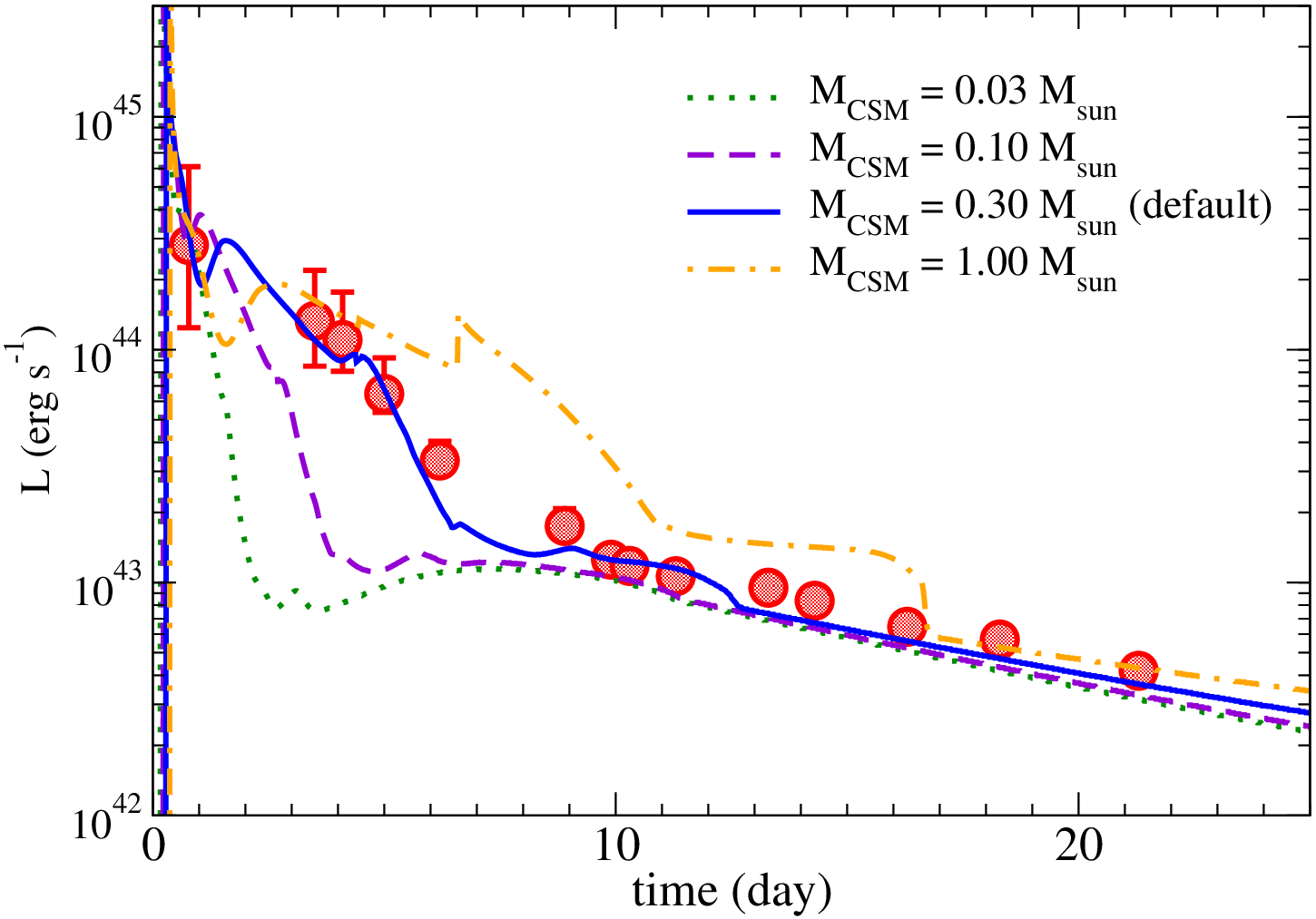}
\includegraphics*[ height=6cm]{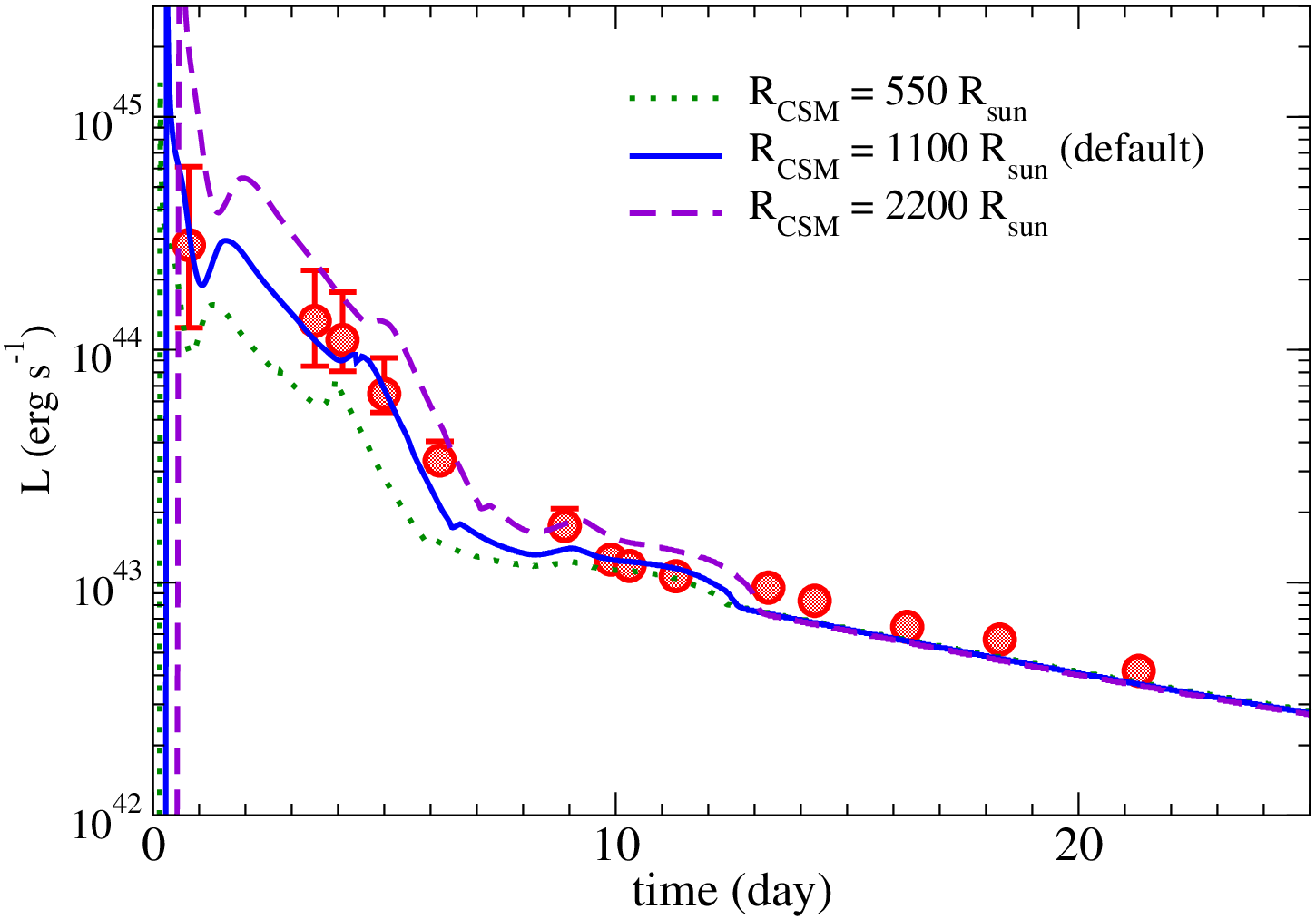}
\caption{
(top panel) Bolometric light curves for the default model (blue solid line) and two comparison models with $5 \times 10^{51}$ erg (green dotted line) and $2 \times 10^{52}$ erg (purple dashed line) respectively.
(middle panel) Same as the top panel but for models including the default model and three comparison models with $M_{\rm CSM} = 0.03~M_{\odot}$  (green dotted line), $0.10~M_{\odot}$ (purple dashed line) and $1.00~M_{\odot}$ (orange dot-dash line). 
(bottom panel) Same as the top panel but for models including the default model and comparison models with $R_{\rm CSM} = 550~R_{\odot}$  (green dotted line) and $2200~R_{\odot}$ (purple dashed line).
}
\label{fig:LC_Eexp}
\end{figure}

The first parameter we examine is the explosion energy.
Because SN 2018gep was a broad-lined Ic SN, the explosion was likely powered by energy sources such as a rapidly rotating magnetar or accretion onto a black hole, which can 
power an explosion with an energy of order $10^{52}$ erg.
In Figure \ref{fig:LC_Eexp}, we plot the SN light curve for models with different explosion energies. The explosion energy has greater importance in the shock-cooling phase (before day 6), where we expect the emergent luminosity to scale linearly with the explosion energy (see Section \ref{analytic}), $L_p \propto E_{\rm exp}$. It is less important in the radioactive decay phase (after day 6) where the luminosity is determined primarily by the $^{56}$Ni mass.
The duration of the shock cooling phase also matches expectations, scaling approximately as $t_p \propto E_{\rm exp}^{-1/4}$ (Section \ref{analytic}) due to the faster expansion of the ejecta that shortens the photon diffusion time

\subsubsection{Dependence on CSM Mass}

The CSM mass and structure depends on the progenitor's mass loss mechanism, which is discussed in previous sections.
In Figure \ref{fig:LC_Eexp}, we plot light curves for models 
with a few values of $M_{\rm CSM}$, but with the same CSM radius and density profile. 
We see that when $M_{\rm CSM}$ decreases, the plateau of the light curve remains at nearly constant luminosity but becomes narrower, with duration scaling approximately as $t_p \sim t_d \propto M_{\rm CSM}^{1/2}$ (Section \ref{analytic}). This occurs because the shock-heated CSM loses its energy more rapidly due to the shorter photon diffusion time of a lower-mass CSM.
All four light curves become similar beyond day 20, where the core $^{56}$Ni-decay dominates the energy production process. 
As discussed above, small CSM masses of $\lesssim \! 0.1~M_{\odot}$ struggle to reproduce the light curve because they produce far too short of a plateau, at least for the CSM structure and radius of our default model.

\subsubsection{Dependence on CSM Radius}

Another important parameter is the CSM radius $R_{\rm CSM}$,
which depends primarily on the CSM expansion velocity and the time delay between the final mass loss and the explosion. 
In Figure \ref{fig:LC_Eexp}, we plot the light curves for models with half and double the CSM radius of the default model.
We see that the CSM radius also strongly affects the early time light curve, with larger CSM radii producing brighter plateaus that scale approximately as $L_p \propto R_{\rm CSM}$ (Section \ref{analytic}). Larger CSM radii also translate to slightly longer plateau times $t_p$, even though this is not clearly predicted from analytic models. 

\subsubsection{Dependence on $^{56}$Ni Mass}

\begin{figure}
\centering
\includegraphics*[ height=6cm]{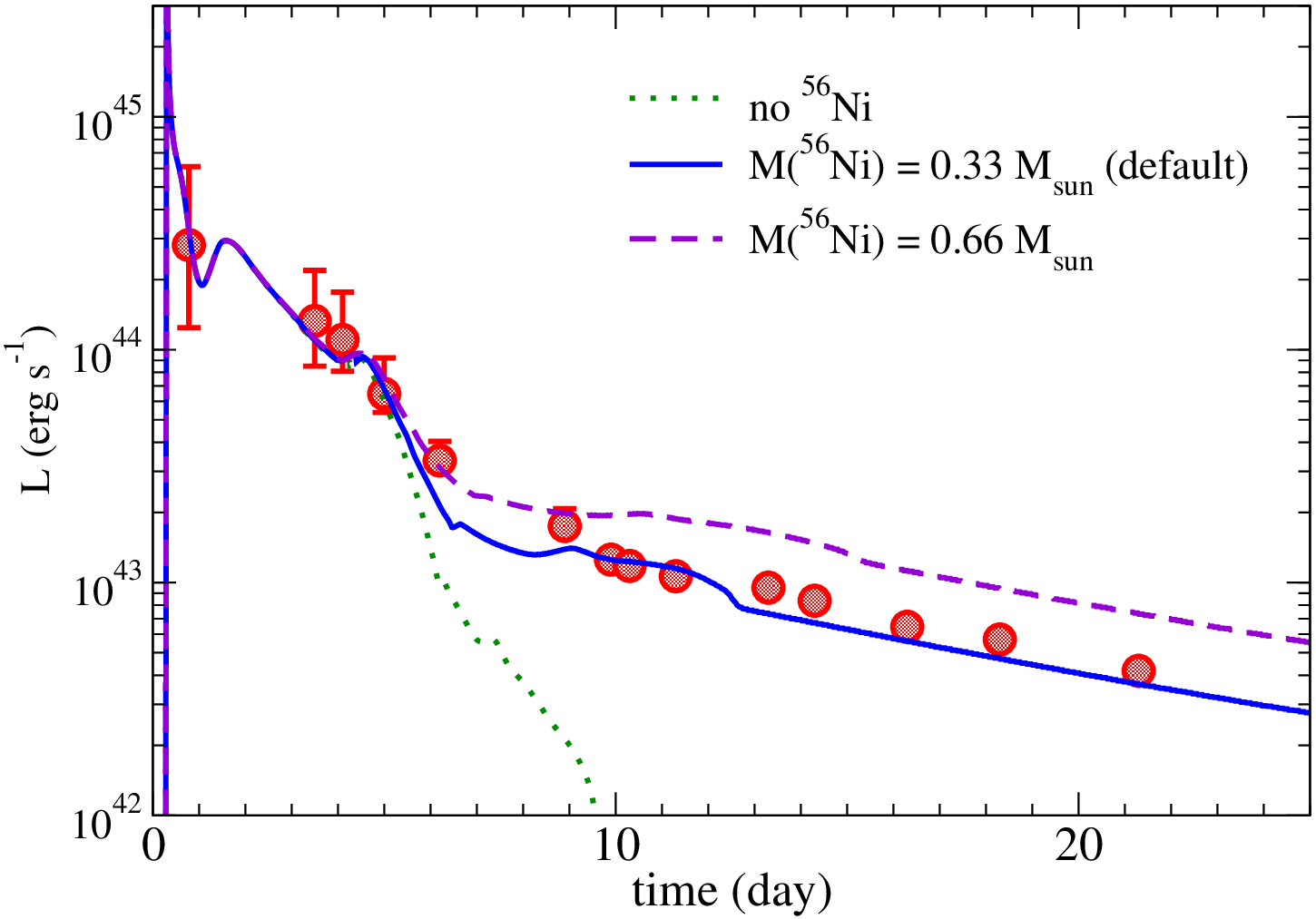}
\includegraphics*[ height=6cm]{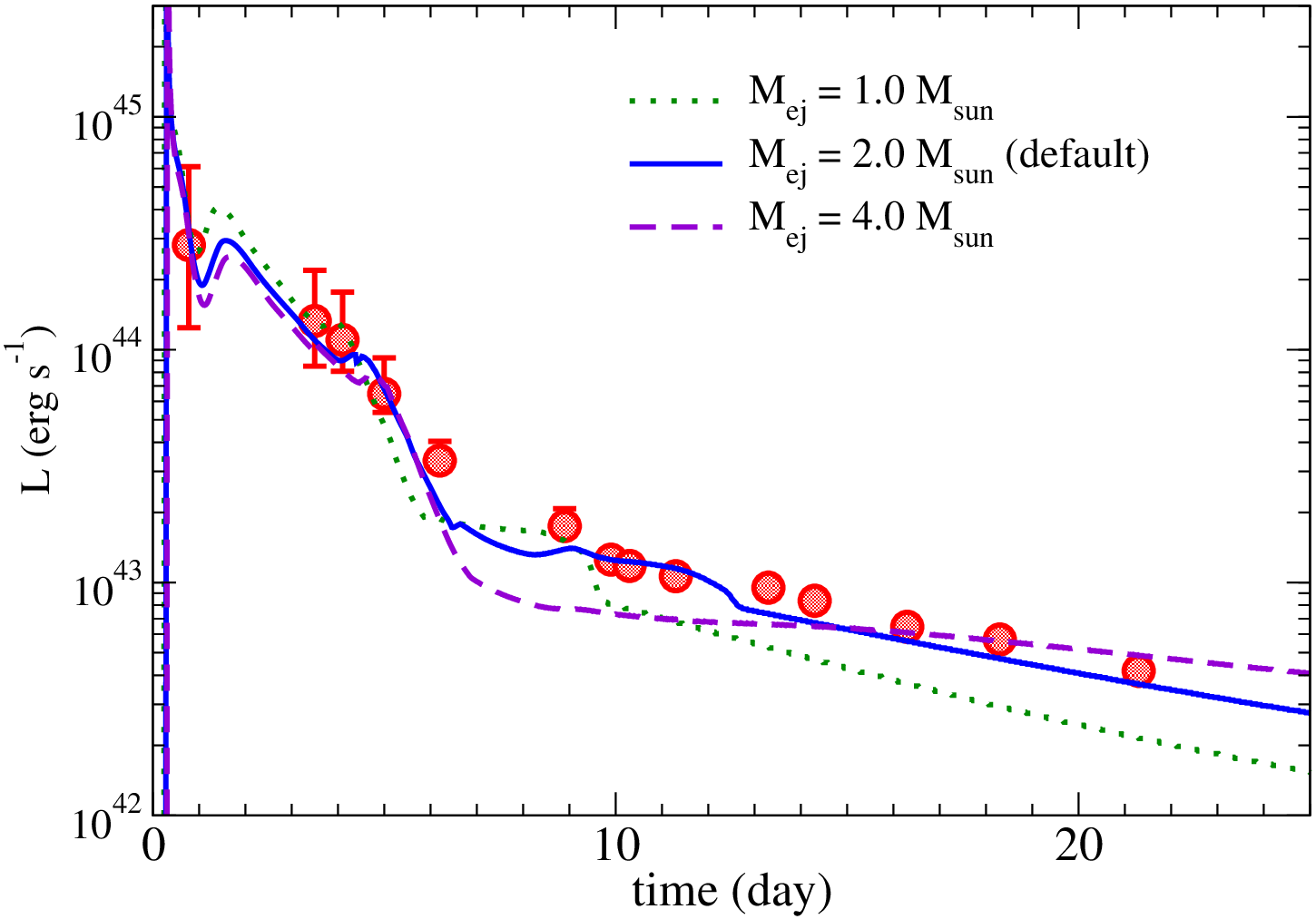}
\includegraphics*[ height=6cm]{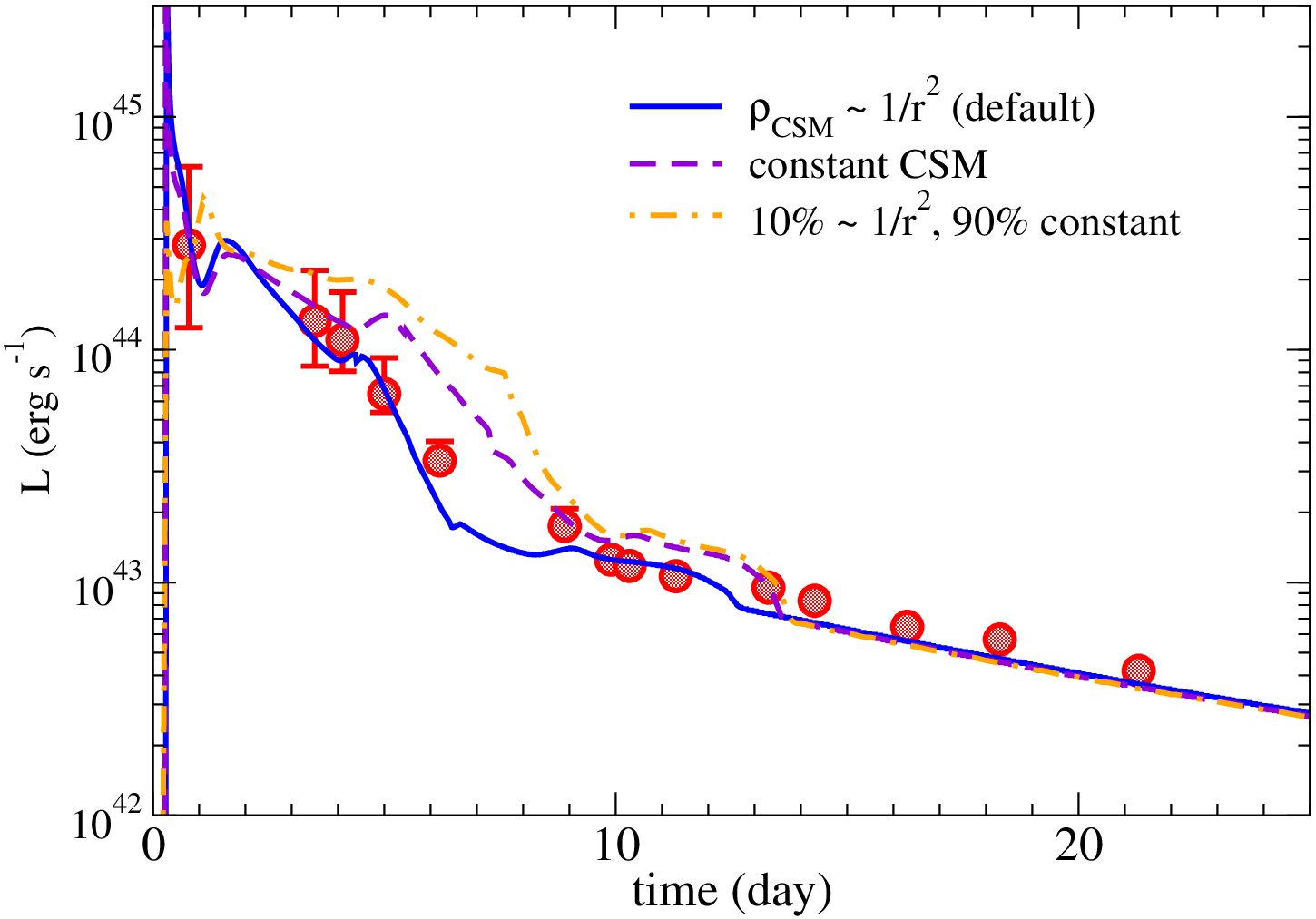}
\caption{
(top panel) Bolometric light curves for the default model (blue solid line) and two comparison models without
(green dotted line) and with 0.66 $M_{\odot}$ (purple dashed line) radioactive $^{56}$Ni. 
(middle panel) Same as the top panel but for models including the default model and two comparison models with ejecta mass $M_{\rm ej} = 1.0~M_{\odot}$ (green dotted line) and 4.0 $M_{\odot}$ (purple dashed line).
(bottom panel) Same as the top panel but for models including the default model and contrasting models with constant CSM density (purple dashed line) and a thin CSM shell contributing 90 \% of $M_{\rm CSM}$ (orange dot-dash line), each with the same outer CSM radius.}
\label{fig:LC_MNi}
\end{figure}

We next examine how the $^{56}$Ni mass affects the 
light curve structure. 
In Figure \ref{fig:LC_MNi}, we compare the default light curve with models containing no $^{56}$Ni or two times more. The zero $^{56}$Ni model is identical to that in Figure \ref{fig:benchmark_model}. Again, the early time light curves are essentially identical in all three cases, demonstrating that the $^{56}$Ni and $^{56}$Co decay do not contribute appreciably to the early time evolution of the light curve. On the other hand, the amount of $^{56}$Ni determines the luminosity of the late-time light curve, similar to normal Type Ib/c SNe.
%Because the expected PPISN progenitors do not lie near the upper boundary of the PPISN regime, the $^{56}$Ni modeled is more likely to be produced by final explosion, instead of being from the pulsation. 
Since the CSM mass and $^{56}$Ni mass are both $\sim 0.3~M_{\odot}$, it is unlikely that the $^{56}$Ni comes from pulsation, and is more likely to be generated during core-collapse. %However, in the extreme case where (jet-like) aspherical pulsation occurs, it is still possible for such overturn of core material to take place.

\subsubsection{Dependence on Ejecta Mass}
\label{sec:LC_ejecta_mass}

The total ejecta mass also affects the light curve. 
In Figure \ref{fig:LC_MNi}, we compare the default models with  two other choices of $M_{\rm ej}$.
For a lower ejecta mass but the same explosion energy, the ejecta has a larger velocity and becomes optically thin sooner. At the low ejecta masses and high explosion energies considered here, the photon diffusion time is short, such that the width and luminosity of the Ni-powered portion of the light curve is not very sensitive to the ejecta mass, as long as the Ni-decay power is thermalized (which is assumed by our SNEC models). However, we still see that higher ejecta masses stay optically thick for longer, leading to a slower decline at late times relative to low ejecta masses.

\subsubsection{Dependence on CSM Density Profile}
\label{sec:LC_shape}

Lastly, we examine how the light curve depends on the slope of the CSM.
In Figure \ref{fig:LC_MNi} we plot the light curve of our default model with a contrasting model in which $\rho_{\rm CSM}$ is constant, but with the same CSM mass and outer radius. We observe that the constant CSM model creates a slightly more extended plateau during the shock cooling phase and the drop of luminosity is slightly slower during the transition towards the $^{56}$Ni-decay phase. 
We also computed a model where most of the CSM is in a thin shell, such that the inner CSM has $\rho_{\rm CSM} \sim 1/r^2$ and the outer CSM has constant $\rho_{\rm CSM}$. The transition takes place at $R_{\rm trans} = 0.9 R_{\rm CSM}$ and the the outer shell has $M_{\rm shell} = 0.9 M_{\rm CSM}$, producing a density contrast of $\sim \! 1000$ across the transition. The thin shell, being more compact, slightly delays the recession of the photosphere relative to a constant $\rho_{\rm CSM}$. Beyond day 15, the three models are identical. We conclude that the CSM mass may be slightly smaller than our default model if it is distributed with constant density or in a thin shell.

\section{Discussion}
\label{sec:discussion}

\subsection{Comparison with Models in the Literature}

Only a couple detailed models have been published for SN 2018gep. Based on the sharp rise and high surface temperature, \cite{Ho2019} argued that an extended shock breakout in a circumstellar medium (CSM) powers the initial peak. Their gray-opacity radiative transfer calculation based on analytic CSM profiles found the light curve could be approximately reproduced with $M_{\rm ej} = 8~M_{\odot}$, explosion energy $E_{\rm ej} = 2 \times 10^{52}$ erg, along with a CSM in a shell with mass $M_{\rm CSM} = 0.02 ~M_{\odot}$ and radius $R_{\rm CSM}$ = $4000 \, R_\odot$ with a small width $400 \, R_\odot$. These numbers indicate the possibility that this compact SN Ic-BL has experienced mass loss very shortly before the explosion. The model of \cite{Ho2019} is distinctive from our default model above, as they predict a higher $R_{\rm CSM}$ but a lower $M_{\rm CSM}$. While their explosion energy is also larger, the energy per unit mass is similar, as they find $E_{\rm exp}/M = 2.5 \times 10^{51}$ erg/$M_{\odot}$ and we find $E_{\rm exp}/M = 3.3 \times 10^{51}$ erg/$M_{\odot}$.

\begin{figure}
\centering
\includegraphics*[height=6cm]{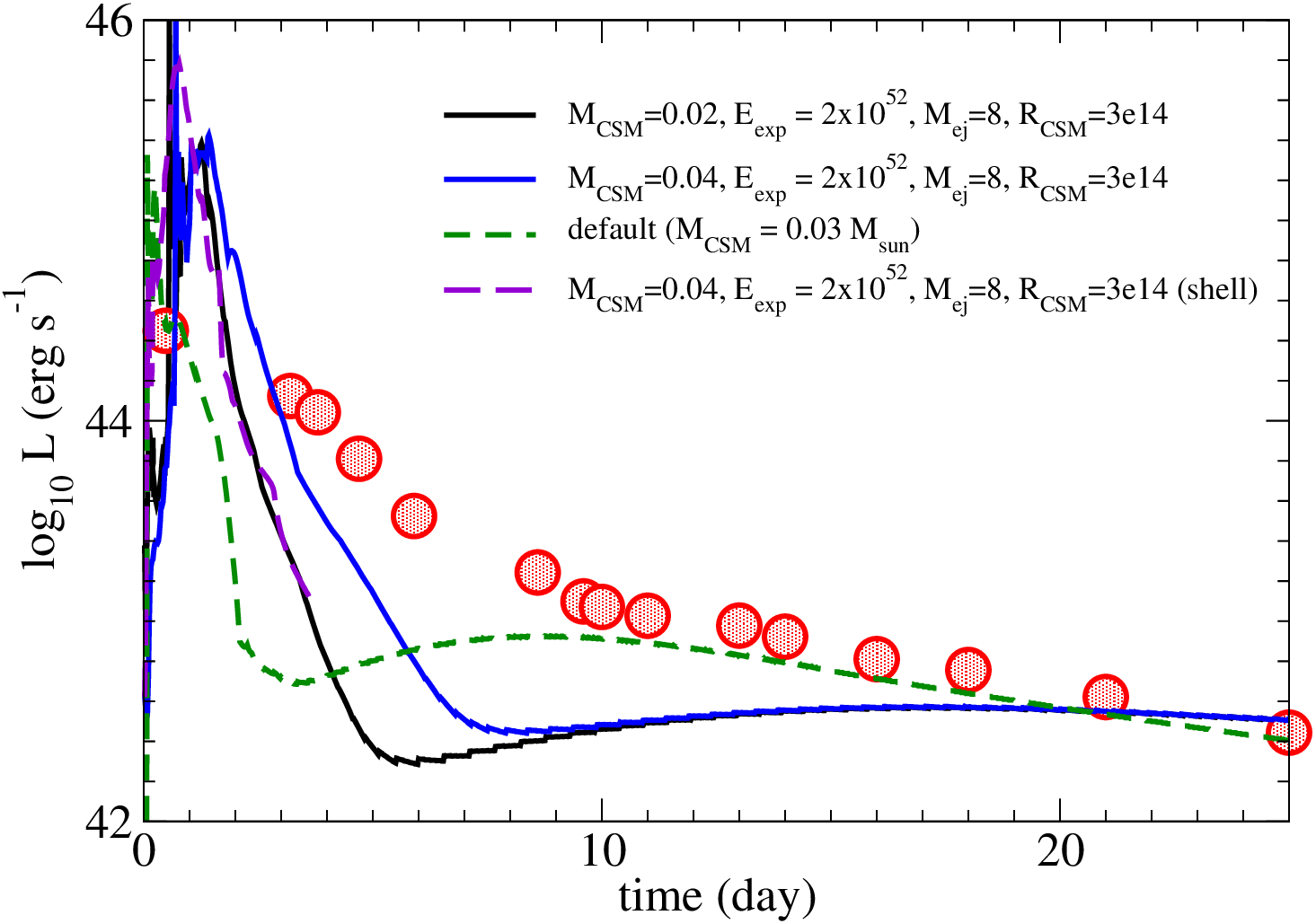}
\caption{
The bolometric light curves predicted by SNEC using the numerical values provided in \citep{Ho2019} (black line), an identical model with twice as much CSM (blue line), and a comparison using our default model but with a lower CSM mass of 0.03 $M_{\odot}$ (green dashed line). The observational data for SN 2018gep  is shown as as red circles.}
\label{fig:LC_compare_Ho}
\end{figure}

To attempt to distinguish between the competing models, we repeat the light curve calculation with SNEC, using their values as the model input. We thus choose a higher progenitor mass model ($M_{\rm ini} =$ 50 $M_{\odot}$, with a pre-explosion mass $M_{\rm exp} \sim 10~M_{\odot}$) such that we can allocate $8~M_{\odot}$ of ejecta mass. In Figure \ref{fig:LC_compare_Ho}, we plot the light curve using their derived values as numerical inputs, along with our default model but with a $M_{\rm CSM}$ similar to theirs. After sharp early light curve peaks, all of these models fade much more rapidly than the data because the low-mass CSM becomes optically thin very quickly.
The luminosity plateaus after day 10, when $^{56}$Ni decay dominates the energy source.  We also repeat the experiment with a larger CSM mass $M_{\rm CSM} = 0.04~M_{\odot}$, but the light curve still fades too fast.

We suspect the main difference (see discussion below) is that the constant opacity used in \cite{Ho2019} is larger than the temperature-dependent opacity in our models. The use of a constant opacity allows the CSM to remain optically thick even after recombination, which slows down the luminosity drop in an unphysical manner. Another difference between their modeling and ours is the use of a thin shell of CSM 
in \cite{Ho2019}, 
which creates a higher CSM density for a given CSM mass, thus allowing the high luminosity to sustain for a longer time. These model differences also change the interpretation the data between day 1 -- 4. 
In our models, shock cooling of the envelope creates a plateau in the first few days, while the extended shock breakout models of \cite{Ho2019} create a large spike in between the data points at 1 and 3 days. Available photometry during that time seems to indicate a luminosity plateau rather than a spike, favoring our shock cooling model. However, the lack of UV data prevents an accurate bolometric luminosity measurement, so we cannot confidently exclude the extended shock breakout model.

To get a rough estimate of the explosion properties of SN 2018gep, \cite{Pritchard2020} compared multi-color light curves to semi-analytic MOSFiT models \citep{guillochon:18}. In their best-fit CSM-interaction model, they derive an ejecta mass 0.49 $M_{\odot}$, $^{56}$Ni mass of 0.13 $M_{\odot}$, CSM mass of $0.11~M_{\odot}$, and CSM radius of $3000 \, R_\odot$. Their inferred CSM mass is in between our shock cooling model (0.3 $M_{\odot}$) and the shock breakout model (0.02 $M_{\odot}$) from \cite{Ho2019}, and their CSM radius is closer to that of \cite{Ho2019}.
However, the best-fit constant opacity of $\kappa = 0.58 \, {\rm cm}^2{\rm g}^{-1}$ for their CSM models is likely unphysically large based on the opacities from our more realistic models (Figure \ref{fig:benchmark_hydro}). This problem is even more pronounced for their best-fit magnetar models, so magnetar powering of SN 2018gep is unlikely.
For these reasons, we believe the parameters of our more detailed CSM-interaction models are closer to the actual characteristics of SN 2018gep. However, a somewhat larger CSM radius and smaller CSM mass than our default model is certainly possible given degeneracies in light curve fitting (see Section \ref{degeneracies}).

\subsection{Comparison with Analytic Model}
\label{analytic}

For an early light curve is dominated by shock cooling of the extended CSM, our models reveal clear trends in terms of light curve plateau luminosity, $L_p$, duration $t_p$, and slope. Here we compare with the scalings predicted by analytic models \citep{Piro2020}. The asymptotic ejecta speed  \citep{Matzner1999} is 
\begin{equation}
    v_t \sim \left( \frac{E_{\rm exp}}{m(r)} \right)^{1/2} \left( \frac{m(r)}{\rho r^3} \right)^\beta \, ,
\end{equation}
with $m(r)$ the enclosed mass, and $\beta \simeq 0.19$. Our models are characterized by a radially extended but low-mass CSM with $M_{\rm CSM} \ll M_{\rm ej}$. Hence, within the CSM, $m(r) \simeq M_{\rm ej}$ and $\rho r^3 \sim M_{\rm CSM}$, such that the CSM ejecta velocity is roughly
\begin{equation}
    v_t \sim E_{\rm exp}^{1/2} M_{\rm ej}^{\beta-1/2} M_{\rm CSM}^{-\beta} \, .
\end{equation}

The duration of the shock cooling light curve is determined by the diffusion timescale $t_d$ of photons out of the shocked CSM,
\begin{equation}
    t_d \sim \left[ \frac{\kappa M_{\rm CSM}}{ v_t c} \right]^{1/2},
\end{equation}
where $\kappa$ is the opacity.
%$K \sim 0.12$ is a constant derived from mass conservation, $n$ is the radial dependence of the CSM profile ($\rho_{\rm CSM} \sim r^n$).
At early times ($t \lesssim t_d$), the luminosity evolves as 
\begin{equation}
    L(t) \sim \frac{c R_{\rm CSM} v_t^2}{\kappa} \bigg(\frac{t_d}{t}\bigg)^{4/(n-2)} \, ,
\end{equation}
The luminosity and photospheric radius fall sharply after a time scale 
\begin{equation}
    t_{\rm ph} \sim \left[ \frac{\kappa M_{\rm CSM}}{ v_t^2} \right]^{1/2} \, .
\end{equation}

In the constant opacity approximation, the plateau luminosity is therefore expected to scale approximately as $L_{p} \propto R_{\rm CSM} E_{\rm exp} M_{\rm CSM}^{-2 \beta} M_{\rm ej}^{2 \beta-1}$. The light curve will drop sharply after a plateau duration $t_{\rm ph} \propto E_{\rm exp}^{-1/2} M_{\rm CSM}^{1/2+\beta} M_{\rm ej}^{1/2-\beta}$. Hence, for a constant plateau luminosity and duration, we expect $R_{\rm CSM} \propto M_{\rm CSM}^{-1}$ and $E_{\rm exp} \propto M_{\rm CSM}^{1 + 2 \beta} M_{\rm ej}^{1 - 2 \beta}$. The scalings provided above are approximately consistent with the light curve modeling results of the previous sections. 

\begin{figure*}
\centering
\includegraphics*[ height=6cm]{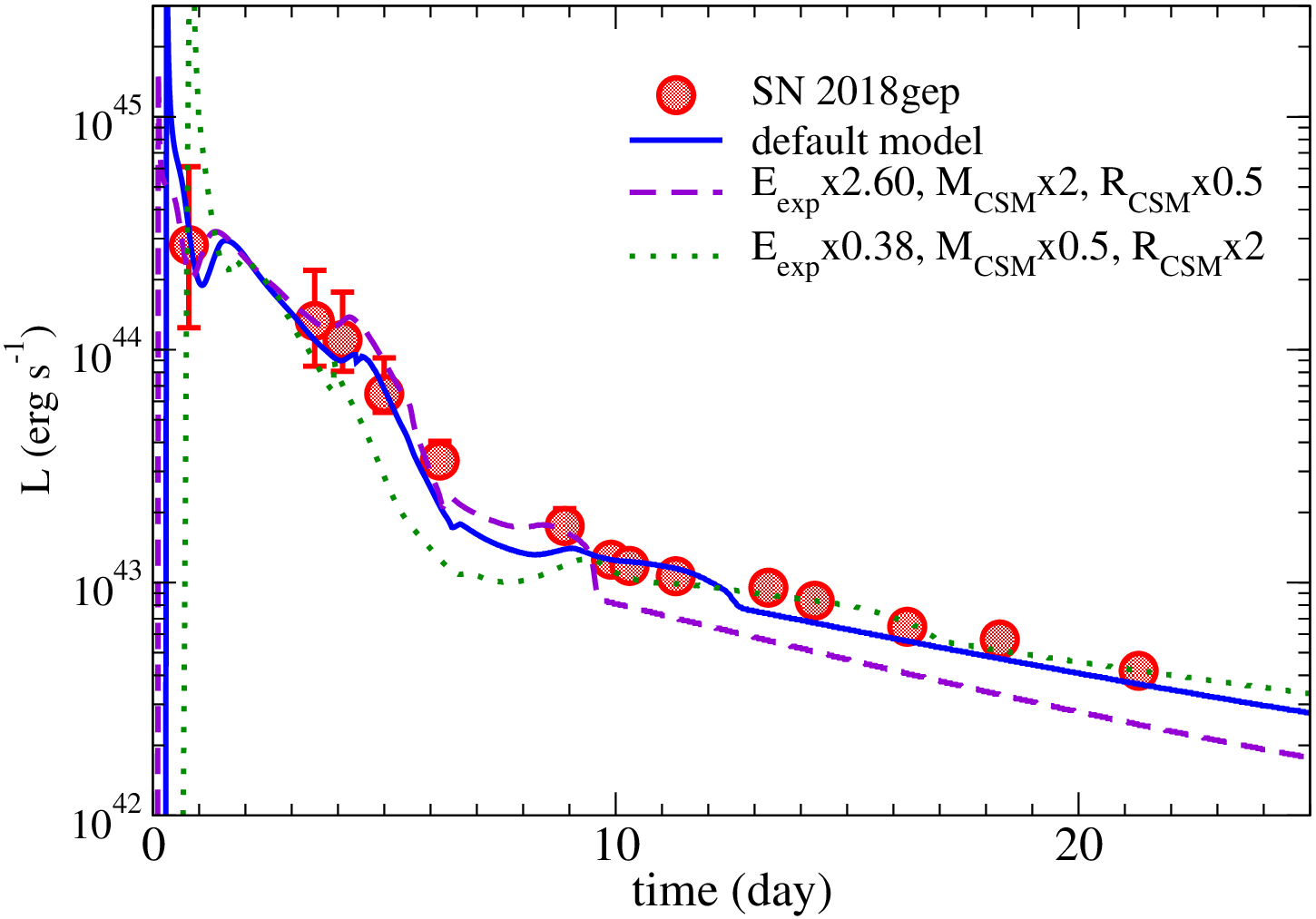}
\includegraphics*[ height=6cm]{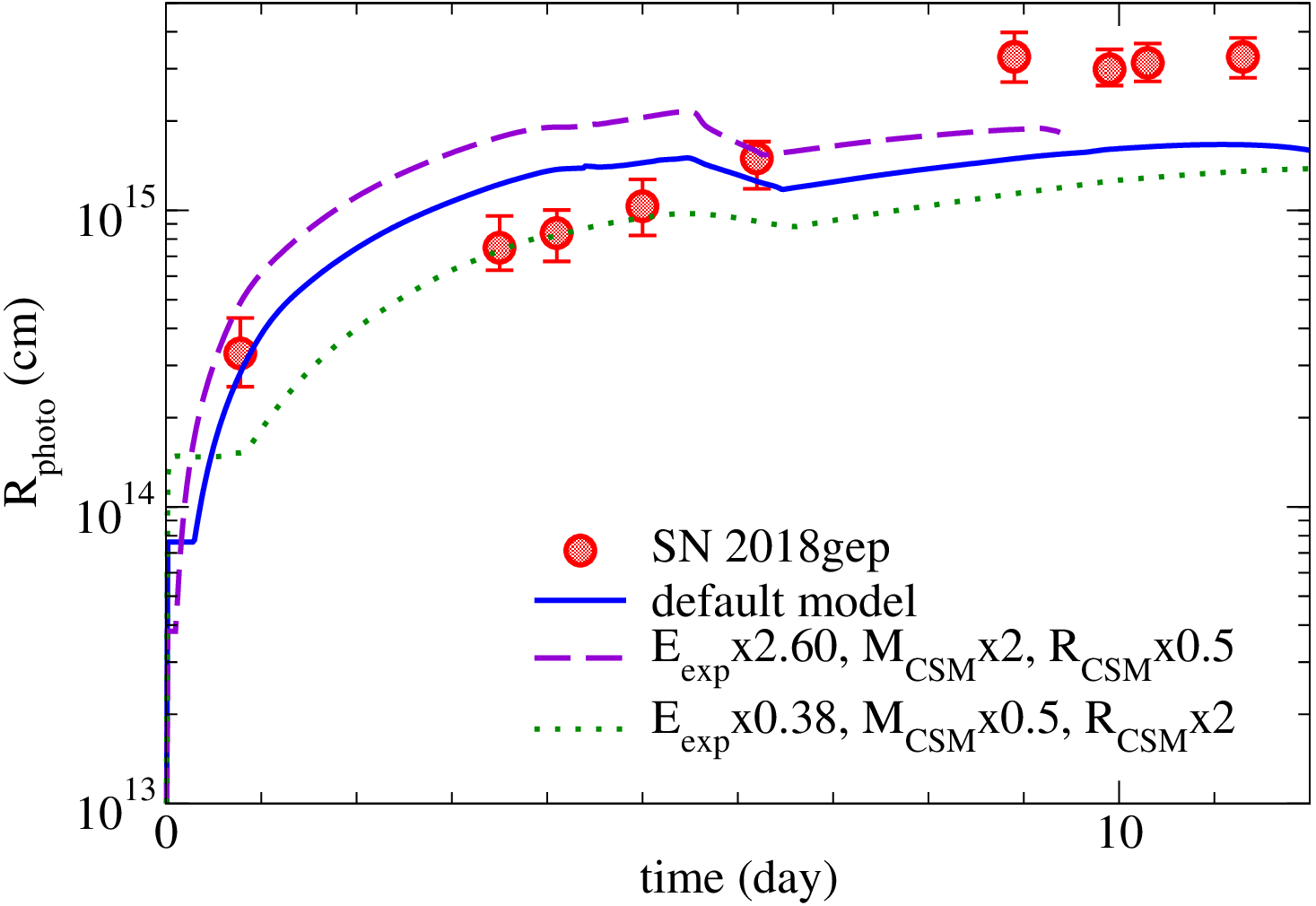}
\includegraphics*[ height=6cm]{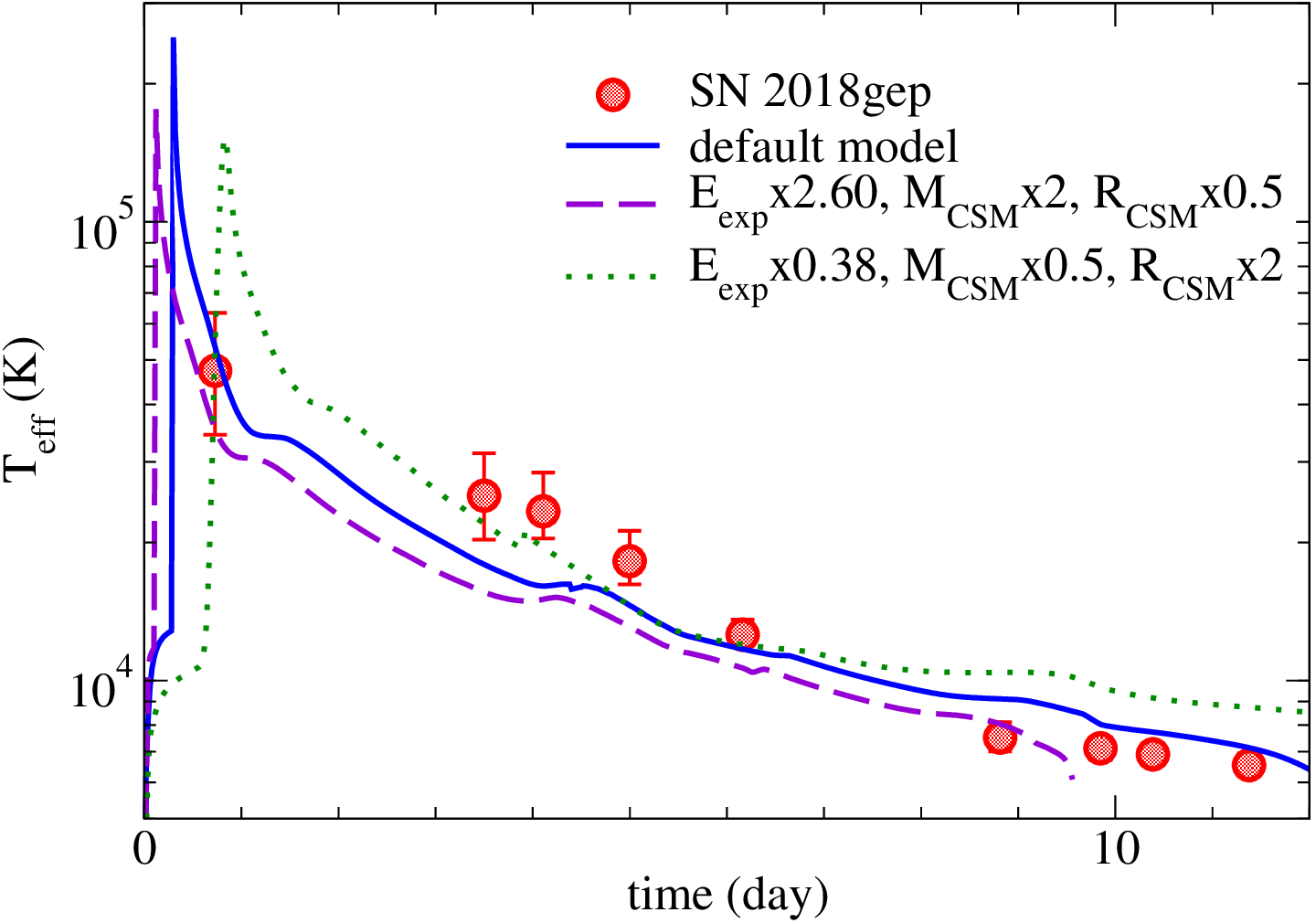}
\includegraphics*[ height=6cm]{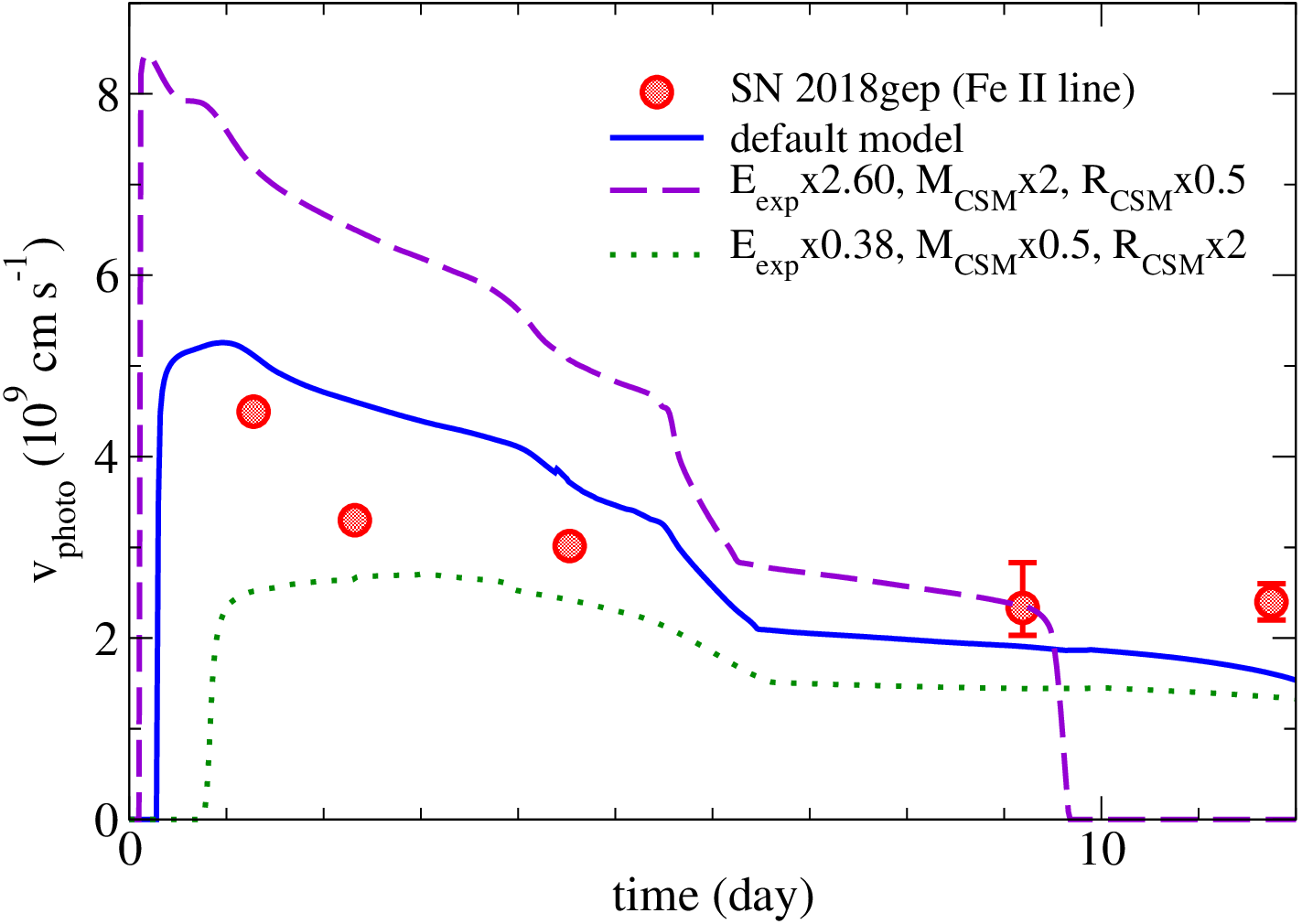}
\caption{(top left) The bolometric light curve of the default model and the contrasting model with $E_{\rm exp} \times 2.60$, $M_{\rm CSM} \times 2.60$ and $R_{\rm CSM} \times 0.5$ (purple dashed line) and $E_{\rm exp} \times 0.38$, $M_{\rm CSM} \times 0.5$ and $R_{\rm CSM} \times 2$ (green dotted line). (top right) Same as top left panel but for the photosphere radius. (bottom left) Same as top left panel but for the effective temperature. (bottom right) Same as top left panel but for the photosphere velocity. 
}
\label{fig:LC_degen}
\end{figure*}

\subsection{Light Curve Degeneracies}
\label{degeneracies}

The above analytic relations for the light curve luminosity and plateau duration predict degeneracies between the CSM mass, radius, and explosion energy, as are well known for Type II-P SNe \citep{Goldberg2020}. To understand whether our inferred CSM parameters for SN 2018gep are subject to such degeneracies, we computed additional numerical models which are analytically predicted to have the same plateau luminosity and duration. In Figure \ref{fig:LC_degen}, we plot the light curves for the default model (solid blue line) and a contrasting model with $2.60 \times E_{\rm exp}$, $2 \times M_{\rm CSM}$ and $0.5 \times R_{\rm CSM}$ (Model 1, purple dashed line), and another with $0.38 \times E_{\rm exp}$, $0.5 \times M_{\rm CSM}$ and $2 \times R_{\rm CSM}$ (Model 2, green dashed line).

We find that the analytic scaling relations work fairly well (though not perfectly) in the parameter range considered, as the numerically computed luminosity, photospheric radius and effective temperature for the three models are similar.
The more compact model (Model 1) has a slightly lower peak luminosity and slower fall in luminosity during the shock-cooling phase. Minor deviations from the analytic predictions occur because the differing density and temperature changes the opacity, which affects how the photosphere recedes. These results echo the findings of \cite{Dessart2019,Goldberg2020} for type-II-P supernova light curves, where they showed that the light curve shape allows for degeneracy between the mass and radius of the progenitor's hydrogen envelope.

However, in contrast to Type II-P SNe where the envelope mass dominates the ejecta mass, the inferred CSM mass for SN 2018gep is much smaller than the total ejecta mass. This means the the photospheric velocity scales approximately as $v \propto E_{\rm exp}/M_{\rm ej}$ and is not plagued by the same degeneracy noted by \cite{Goldberg2020} for Type II-P SNe, so the photospheric velocity can potentially be used to break the degeneracy. In this case, a higher $E_{\rm exp}$ (Model 1) leads to a higher photosphere velocity, which predicts photospheric velocities substantially larger than those observed at early times (Figure \ref{fig:LC_degen}). A better match is obtained for our default model or lower $E_{\rm exp}$ (Model 2). We caution that the measured Fe II line velocity is highly uncertain and is likely to be slightly larger than the continuum photospheric velocity because it is formed farther out in the ejecta (e.g., \citealt{Paxton2018}). 
For example, in iPTF14hls \citep{Arcavi2017}, the photosphere radius derived from the Fe II line exceeds that of the blackbody radius after 100 days, exemplifying that the two radii may be very different.

We conclude that the CSM properties likely lie somewhere in between the default model and Model 2, i.e., in the range $0.15 \, M_\odot \lesssim M_{\rm CSM} \lesssim 0.3 \, M_\odot$, $1100 \, R_\odot \lesssim R_{\rm CSM} \lesssim 2200 \, R_\odot$, and $5 \times 10^{51} \, {\rm erg} \lesssim E_{\rm exp} \lesssim 10^{52} \, {\rm erg}$. More detailed modeling should be performed to identify other possible degeneracies (e.g., between $E_{\rm exp}$ and $M_{\rm ej}$, whether these can be broken by the Ni-powered portion of the light curve, and to robustly estimate uncertainties in the inferred explosion parameters). 
In the future, it will be also beneficial to use  realistic radiative transfer models during the mass ejection process so that the exact density profile of the CSM can be accurately determined. Spectral synthesis which distinguishes velocities of individual elements will also provide an alternative method to lift the degeneracy.

\subsection{Mixing and Supernova Type}

\cite{dessart:20} showed that a star with no H-rich envelope explodes as a Type Ib (Ic) supernova depending on its $X$(He) at the photosphere. According to Table 3 and Figure 3 in that work, a star with a photospheric He mass fraction $X$(He) $\lesssim 0.2$ explodes as a Type Ic and $X$(He) $\gtrsim
0.5$ as a Type Ib supernova.  They also mentioned that $^{56}$Ni
mixing is less important because of the long mean free path of
gamma-rays. Our default model of $M_{\rm ini} = 25 ~M_\odot$ contains $X$(He) $\sim 0.36$ (Figure \ref{fig:init_abund}) and thus lies at the transition between Ib and Ic SNe. In higher mass models in Section \ref{sec:LC_shape}, $X$(He) = 0.16, 0.037, and 0.035 after mixing for $M_{\rm ini} =$ 40, 50, and $60~M_{\odot}$, respectively. These models would appear as Type Ic SNe.

It is necessary to conduct detailed radiation hydrodynamical simulations to reliably determine the SN spectral type. Additionally, the mixing is likely non-spherical, so that $^{56}$Ni and He may not be microscopically mixed. For example, for a high-energy jet-induced explosion \citep{Nomoto2017H}, much of the ejected mass that powers the light curve may originate near the CO core
where energy deposition takes place. Such jet-associated ejecta and fallback materials effectively reproduce mixing (Figure 1 of \cite{Tominaga2007}),
and in this case we expect Type Ic spectra.

\subsection{Asphericity and Jet-like Explosion}
\label{sec:aspherical} 

The relatively low ejecta mass of our default model is very small compared to the progenitor mass for PPISNe (i.e. progenitor masses $M_{\rm ini}$ of roughly 80 -- 140 with pre-explosion masses $M_{\rm exp}$ of roughly 35 -- 50 $M_{\odot}$). Hence, adopting the PPISN interpretation likely requires that the ejecta be highly aspherical (e.g., a bipolar jet-driven outflow) or that nearly all of the star collapses into a black hole and only a small fraction is ejected \citep[see e.g., ][for aspherical mass ejection in the neutron star case.]{Wongwathanarat2013}  \cite{Powell2021} examines core-collapse and the (partial) explosion of PPISN models. For their low-mass PPISN model, a successful explosion of $\sim 3 \times 10^{51}$ erg is observed but the high binding energy of the remnant implies black hole formation with partial mass ejection. This phenomenon could account for the low ejecta mass found by our light curve modeling. The escape of $^{56}$Ni into the ejecta could be produced by this partial explosion, or in a jet or BH accretion disk wind.
%See Powell et al.(2021) for a recent exploration on the mass ejection for this class of object.

Both of these possibilities may be realized in the collapsar scenario for driving the explosion, in which the explosion energy arises from a jet and/or disc wind from an accreting black hole \citep{Woosley2006, Woosley2006b}. Numerical simulations of collapsars show that the conservation of angular momentum leads to an accretion disk with a standing shock \citep[e.g.][]{Diego1996, Stone1999}. The disk can drive an outflow containing a couple solar masses, a few tenths of a solar mass of $^{56}$Ni, and an energy of $\sim \! 10^{52} \, {\rm erg}$ (see e.g., \citealt{macfadyen:99,kohri:05,zenati:20}), similar to the properties of our default model. Alternatively, the explosion could be driven by a rapidly rotating magnetar (e.g., \citealt{obergaulinger:20}).
%The instability of the accretion disk due to gravitational and magneto-rotational instabilities can trigger a sudden outburst of energy in the form of jet \citep{Tsuruta2019}.
The jet \citep[e.g.,][]{Maeda2002,Nagataki2003, Nomoto2013} can cause shock heating in the envelope, which creates a distinctive nucleosynthetic pattern including lots of $^{56}$Ni \citep{Tominaga2009}. The breakout can lead to a long gamma-ray burst \citep{Tominaga2007, Nagataki2011}. However, whether the jet breaks out depends strongly on its energetics \citep{Zhang2008, Barnes2018}, with a smothered jet resulting in a Type Ic-BL event like SN 2018gep. 

One-dimension models of massive stellar explosions \citep[e.g.,][]{Heger2010, Nomoto2017H,  Limongi2020}, predict that an explosion energy of
roughly $2 \times 10^{51}$ erg is necessary to produce $\sim \! 0.3~M_{\odot}$ of $^{56}$Ni in the ejecta, for progenitor models with masses $M_{\rm ini} \lesssim 20 \, M_{\odot}$. Higher explosion energies of roughly $5 \times 10^{51}$ erg are required for higher mass stars \citep{Umeda2008}. 
The low-mass progenitor possibility is roughly consistent with our wave-driven models (though the explosion energy is lower than that inferred from light curve modeling), while the high-mass possibility is more consistent with our PPISN models. In the low-mass progenitor scenario, the small ejecta mass inferred from light curve modeling is expected from the $\sim \! 4-5 \, M_\odot$ helium core of the progenitor star.

In the high-mass PPISN progenitor scenario, a small ejecta mass requires asymmetric explosion models in which most of the $^{56}$Ni and explosion energy is contained within a small amount of mass, possibly ejected within bipolar jets. In this case, the $^{56}$Ni considered in our light curve models comes from the final explosion (and not from prior pulses). It is possible that, for strong pulsations, a significant amount of $^{56}$Ni is synthesized prior to the final explosion. The radioactive energy can thermalize in the core and lead to further mass loss. However, this usually happens for high-mass PPISN near the pair-instability supernova limit, whereas our light curve modeling suggests lower mass PPISN progenitors.

It will be interesting future work to self-consistently model the pre-SN evolution of a rapidly rotating helium star with pre-SN mass of $\approx 40 \, M_\odot$. If it can produce pre-SN ejecta of $\sim \! 0.3~M_{\odot}$ due to pulsational pair instability before collapse, and generate a $\sim \! 10^{52} \, {\rm erg}$ explosion with a few solar masses of high-velocity ejecta, it could likely produce an event like SN 2018gep. 
Recent work \citep{Marchant2020} shows that whether pulsation occurs depends on how fast the star is rotating.  Their rigidly rotating models show pulsations which release less than half of the stellar angular momentum. This favors the formation of a collapsar which allows jet-energy deposition for aspherical mass ejection. Future work should further examine the interplay of rotation, stellar evolution, mass ejection, and the supernova explosion.

\subsection{Caveats}

In this work we have relied on the stellar evolution code MESA for modeling the mass ejection and the formation of CSM before the final explosion. MESA assumes diffusive radiation transport, and it assumes that matter is in local thermal equilibrium (LTE) with radiation. These approximations will gradually break down as the density of the ejected matter drops as it expands. The transition from optically thick to optically thin matter allows the radiation to leak out more efficiently, hence lowering the radiation pressure driving expansion.
The exact temperature profile and opacity profile will also depend on this radiation transport, further complicating the problem, and potentially altering the density profile away from that of our models.

However, we do not expect the CSM mass or radius to be strongly altered by more accurate radiative transport calculations. As shown in the wave-driven mass loss model in Figure 4 of \cite{Fuller2018}, the ejected material has nearly reached its asymptotic velocity by the time that radiation diffuses effectively, so the expansion velocity calculated from MESA is a reasonable estimate. As long as the ejecta mass is optically thick where it is being accelerated (i.e., near the sonic point and escape point), the wave-driven ejecta speed tends to be near or somewhat larger than the escape speed, as described in \cite{Quataert2016}. We suspect PPISN mass loss to also be driven in the optically thick limit so that the calculation of \cite{Renzo2020} is a reasonable approximation.
%Our model here is then an optimistic estimation to the extent of CSM produced by the wave-driven model (with similar argument applicable to the ejected matter in PPISN model).
For accurate CSM modeling out to lower optical depths, multi-band radiative transfer alongside hydrodynamics will be necessary. Non-radial instabilities between colliding shells will also affect the density profile \citep{Chen2014}, so multi-dimensional calculations should also be examined. However, this will greatly increase the required computational resource, so we leave these investigations for future work.

In our radiative transfer modeling, we approximated the CSM density profile with a $1/r^2$ dependence. As shown in Figures \ref{fig:profile_tdep} - \ref{fig:profile_energy}, this scaling captures the global CSM features. However, a fully consistent model will require a more realistic density profile which also contains density jumps at shell-shell collisions. Our exploratory model of a thin and dense CSM shown in Section \ref{sec:LC_shape} suggests that some quantitative features of the CSM will change slightly with the CSM internal structure. It will be an interesting follow-up project to understand how these features affect the SN light curve.

\section{Conclusion}

We have simulated the process of wave-driven pre-supernova mass loss in hydrogen-free supernova progenitors, examining the mass loss as a function of the progenitor's pre-explosion mass, the total deposited energy, and the deposition duration. Within the parameter range motivated by detailed stellar evolutionary models, a larger CSM mass can result from a shorter deposition duration (for a fixed energy) or a higher energy deposition (for a fixed duration). Low pre-explosion progenitor masses ($M_{\rm exp} \lesssim 10 \, M_\odot$) appear to be necessary for producing large CSM masses in the range $M_{\rm CSM} \gtrsim 10^{-2}~M_{\odot}$, because the specific binding energy is much larger for higher mass helium stars, so the energy budget of the wave heating mechanism is likely not sufficient to eject large amounts of mass.

We also explored the physical origin of the super-luminous rapid transient SN 2018gep \citep{Ho2019}, spectroscopically classified as a Type Ic-BL SN. The rapid rise and high peak luminosity is best explained via CSM interaction and shock cooling of the CSM (as opposed to extended shock breakout), providing a constraint on the mass-loss mechanism of the progenitor star prior to its final explosion. By modeling the bolometric light curve, we find a good fit for CSM interaction models with explosion energy $E_{\rm exp} = 1\times10^{52}$ erg, CSM mass $M_{\rm CSM} = 0.30 ~M_{\odot}$, CSM radius $R_{\rm CSM} = 1100~R_{\odot}$, ejecta mass $M_{\rm ej} = 2.0~M_{\odot}$, and nickel mass $M(^{56}{\rm Ni}) = 0.33 \, M_\odot$. In this interpretation, the bright early light curve of SN 2018gep is created by shock cooling of the extended CSM, after which the transient evolves into a ``normal" nickel-powered Ic-BL SN because of such high explosion energy as $E_{\rm exp} = 1\times10^{52}$ erg as first shown by the synthetic spectra for the hypernova model by \cite{Iwamoto1998}.

We have extensively tested how the bolometric light curve in the shock-cooling phase and the $^{56}$Ni-decay phase of our models depends on the model parameters above. Our results indicate that our fitting is relatively robust, given the large number of free parameters in the models. 
%We further analyzed how the bolometric light curve shape depends on a number of model parameters, including CSM mass, radius, distribution and so on, among which we examined the degeneracy of the light curve shape on progenitor models.
Like Type II-P SNe, our CSM shock-cooling models do exhibit moderate degeneracy in the light curve shape (peak luminosity and its width) for a given $E_{\rm exp}$, $M_{\rm ej}$, $R_{\rm CSM}$ and $M_{\rm CSM}$. Hence, there could be other combinations of these parameters that provide similarly good fits to the data, so our solution may not be unique.

We show that the interpretation of SN 2018gep could be compatible with two distinct mechanisms: (1) wave-driven mass loss from a fairly low-mass $M_{\rm He} \sim 5\, M_\odot$ helium star progenitor, which ejects $M_{\rm CSM} \sim 0.1 \, M_\odot$ of CSM to large radii, or (2) pulsational pair instability mass loss of a very massive ($M_{\rm He} \sim 40 \, M_\odot$) progenitor, which ejects $M_{\rm CSM} \sim 0.3 \, M_\odot$) in the final weeks of its life. We favor the latter scenario because wave-driven mass loss struggles to eject enough CSM in the shock-cooling interpretation of the light curve. Addtionally, the high-energy broad-lined nature of the SN suggests the explosion is powered by a rapidly rotating central engine unlikely to form in low-mass SN progenitors. This is also consistent with the low-metallicity environment of SN 2018gep, and with its bright observed outburst roughly two weeks before explosion.

In the pulsational mass loss interpretation of SN 2018gep, the progenitor of this event was a $\approx 40 \, M_\odot$ helium star  at the boundary between Type Ic SNe and superluminous Type Ic SNe. Slightly lower mass progenitors do not eject enough mass via pulsational pair instability to affect the light curves, producing Type Ic or Ic-BL SNe. Slightly higher mass progenitors eject much more mass to larger radii, potentially producing longer lasting Type Ic superluminous SNe via CSM interaction. Our interpretation of SN 2018gep may help unify these seemingly distinct classes of SNe into one conceptual framework.

\section{Acknowledgments}

We thank Jared Goldberg, David Khatami, Maryam Modjaz, and the ZTF theory network for useful discussions that helped inspire and refine this work. S.C.L thanks the MESA development community for making the code open-sourced and V. Morozova and her collaborators in providing the SNEC code open source. S.C.L. and JF acknowledges support by NASA grants HST-AR-15021.001-A and 80NSSC18K1017. K.N. has been supported by the World Premier International Research Center Initiative (WPI Initiative), MEXT, Japan, and JSPS KAKENHI Grant Numbers JP17K05382 and JP20K04024.

\software{MESA \citep{Paxton2011,Paxton2013,Paxton2015,Paxton2017,Paxton2018} version 8118 and configuration files \citep{zenodo}, SNEC \citep{Morozova2015}.}

\bibliographystyle{aasjournal}
\pagestyle{plain}
\bibliography{biblio}

%% TABLES
%%
%% If there are any tables, put them here.
%%

\end{document}